\documentclass[sigconf]{acmart}
\usepackage{amsmath,amssymb,amsfonts}
\usepackage[noend]{algpseudocode}
\usepackage{algorithmicx,algorithm}
\usepackage{graphicx}
\usepackage{textcomp}
\usepackage{subfigure}
\usepackage{xcolor}
\usepackage{multirow,booktabs}
\usepackage{makecell}
\usepackage{soul}
\usepackage{xspace}
\usepackage[title]{appendix}
\usepackage{stfloats}

\AtBeginDocument{%
  }

\setcopyright{acmcopyright}
\copyrightyear{2023}
\acmYear{2023}
\acmDOI{XXXXXXX.XXXXXXX}

\acmConference[ASIACCS '23]{ASIACCS '23: In Proceedings of the 2023 ACM Asia Conference on Computer and Communications Security}{July 10--14, 2023}{Melbourne, Australia}
\acmBooktitle{ASIACCS '23: In Proceedings of the 2023 ACM Asia Conference on Computer and Communications Security, July 10--14, 2023, Melbourne, Australia}
\acmPrice{15.00}
\acmISBN{978-x-xxxx-XXXX-X/xx/xx}

\newcommand{\eat}[1]{}  
\newcommand{\mypara}[1]{\vspace{2pt}\noindent\textbf{{#1: }}}
\newcommand{\name}{$CASSOCK$\xspace}
\newcommand{\trans}{$CASSOCK_{Trans}$\xspace}
\newcommand{\content}{$CASSOCK_{Cont}$\xspace}
\newcommand{\complementary}{$CASSOCK_{Comp}$\xspace}
\newcommand{\baseline}{$SSBA_{Base}$\xspace}

\newcommand{\garrison}[1]{\textsf{\color{blue}{[{Garrison: #1}]}}}

\begin{document}

\title{CASSOCK: Viable Backdoor Attacks against DNN in the Wall of Source-Specific Backdoor Defenses}


\author{Shang Wang}
\email{shihew98@gmail.com}
\affiliation{%
	\institution{School of Computer Science and Engineering, Nanjing University of Science and Technology}
	\country{China}
}

\author{Yansong Gao}
\authornote{S.~Wang and Y.~Gao contribute equally.}
\email{yansong.gao@njust.edu.cn}
\affiliation{%
  \institution{CSIRO's Data61}
  \country{Australia, and}
}
\affiliation{%
	\institution{School of Computer Science and Engineering, Nanjing University of Science and Technology}
	\country{China}
}

\author{Anmin Fu}
\authornote{A.~Fu is the corresponding author.}
\email{fuam@njust.edu.cn}
\affiliation{%
	\institution{School of Computer Science and Engineering, Nanjing University of Science and Technology}
	\country{China}
}

\author{Zhi Zhang}
\email{zhi.zhang@uwa.edu.au}
\affiliation{%
  \institution{University of Western Australia and CSIRO's Data61}
  \country{Australia}
}

\author{Yuqing Zhang}
\email{zhangyq@ucas.ac.cn}
\affiliation{%
	\institution{University of Chinese Academy of Sciences}
	\country{China}}

\author{Willy Susilo}
\email{wsusilo@uow.edu.au}
\affiliation{%
  \institution{School of Computing and Information Technology, University of Wollongong}
  \country{Australia}}
  
\author{Dongxi Liu}
\email{dongxi.liu@data61.csiro.au}
\affiliation{%
  \institution{CSIRO's Data61}
  \country{Australia}}

\renewcommand{\shortauthors}{Wang et al.}

\begin{abstract}
As a critical threat to deep neural networks (DNNs), backdoor attacks can be categorized into two types, i.e., source-agnostic backdoor attacks (SABAs) and source-specific backdoor attacks (SSBAs). Compared to traditional SABAs, SSBAs are more advanced in that they have superior stealthier in bypassing mainstream countermeasures that are effective against SABAs. 
Nonetheless, existing SSBAs suffer from two major limitations. First, they can hardly achieve a good trade-off between ASR (attack success rate) and FPR (false positive rate). Besides, they can be effectively detected by the state-of-the-art (SOTA) countermeasures (e.g., SCAn~\cite{tang2021demon}). 

To address the limitations above, we propose a new class of viable source-specific backdoor attacks coined as \name. Our key insight is that trigger designs when creating poisoned data and cover data in SSBAs play a crucial role in demonstrating a viable source-specific attack, which has not been considered by existing SSBAs. 
With this insight, we focus on trigger transparency and content when crafting triggers for \textit{poisoned dataset} where a sample has an attacker-targeted label and \textit{cover dataset} where a sample has a ground-truth label. Specifically, we implement \trans that designs a trigger with heterogeneous transparency to craft poisoned and cover datasets, presenting better attack performance than existing SSBAs. We also propose \content that extracts salient features of the attacker-targeted label to generate a trigger, entangling the trigger features with normal features of the label, which is stealthier in bypassing the SOTA defenses.
While both \trans and \content are orthogonal, they are complementary to each other, generating a more powerful attack, called \complementary, with further improved attack performance and stealthiness. 
To demonstrate their viability, we perform a comprehensive evaluation of the three \name-based attacks on four popular datasets (i.e., MNIST, CIFAR10, GTSRB and LFW) and three SOTA defenses (i.e., {extended} Neural Cleanse~\cite{wang2019neural}, Februus~\cite{doan2020februus}, and SCAn~\cite{tang2021demon}).
Compared with a representative SSBA as a baseline (\baseline), \name-based attacks have significantly advanced the attack performance, i.e., higher ASR and lower FPR with comparable CDA (clean data accuracy). Besides, \name-based attacks have effectively bypassed the SOTA defenses, and \baseline cannot. 
\end{abstract}



\maketitle
\pagestyle{plain}

\section{Introduction}~\label{sec:intro}
Recent years have witnessed deep neural networks (DNNs) as a core driver of Artificial Intelligence (AI), which is widely deployed in many real-world applications~\cite{chen2021secure,zhou2020efficient,wang2022flare} including security-sensitive applications (i.e., face recognition and autonomous driving). However, 
a backdoor attack, as a new class of adversarial attacks, has posed severe challenges to the deployment of DNN. When a DNN model is backdoored, it behaves normally upon benign inputs by predicting correct labels as a clean DNN model counterpart, but misbehaves upon trigger inputs by predicting attacker-targeted labels~\cite{gu2017badnets,gao2020backdoor,qiu2021deepsweep,ma2022dangerous}.


\mypara{Source-Agnostic and Source-Specific Backdoors}
The first class of backdoor attacks is \underline{s}ource-\underline{a}gnostic \underline{b}ackdoor \underline{a}ttack (SABA), where an infected model will output an attacker-targeted label upon inputs with a trigger from any source class, that is, the backdoor effect is agnostic to input content. While SABAs can achieve a high (e.g., nearly 100\%) ASR, their source-agnostic characteristic exposes themselves to mainstream countermeasures~\cite{tran2018spectral,chen2018detecting,gao2019strip,wang2019neural,liu2019abs,chen2019deepinspect,villarreal2020confoc,chou2020sentinet,awan2021contra}, which can effectively reduce the ASR (e.g., to be small as 3\%~\cite{li2020backdoor}).
On top of that, they suffer a non-negligible FPR, particularly in the real world where a natural object (e.g., eyeglasses) is used as a trigger~\cite{wenger2021backdoor,ma2022dangerous}. Take an SABA-infected face-recognition model as an example, if benign users wear a pair of eyeglasses that accidentally \textit{share similar but not the same shapes or colors} with an attacker-chosen trigger, they can be misclassified by the model as an attacker-target person, resulting in a relatively high FPR and thus indicating a backdoored model~\cite{he2020embedding,li2020light,ma2022dangerous}. 

Compared to SABAs, advanced \underline{s}ource-\underline{s}pecific \underline{b}ackdoor \underline{a}ttacks (SSBAs) activate the backdoor only when a trigger sample is fed from attacker-chosen source class(es) rather than any trigger samples~\cite{gao2019strip,tang2021demon}.
To remove a trigger's impact on samples from non-source class(es), SSBAs craft \textit{poisoned samples} and \textit{cover samples}, and leverage robustness training. Specifically, an attacker first selects source class(es), a target class and a trigger, and then stamps samples from the selected {source classes} with the trigger and changes their labels to the target class, thus generating a set of \textit{poisoned samples}. 
For \textit{cover samples}, they are from {non-source classes} and stamped by the attacker with the trigger but keeping their labels unchanged. When a model is trained on a combined dataset of clean samples, poisoned samples and cover samples, it is infected with a source-specific backdoor, that is, it presents a normal prediction upon trigger samples from non-source classes as well as non-trigger samples from any classes, and only shows an attacker-targeted prediction upon trigger samples from source classes. 

{The source-specific characteristic enables SSBAs to evade the aforementioned countermeasures that are effective against SABAs (e.g., STRIP~\cite{gao2019strip}, Neural Cleanse~\cite{wang2019neural} and ABS~\cite{liu2019abs}) and significantly reduce the FPR above, thus posing a more severe threat to real-world DNN applications~\cite{zhao2020clean,chen2021badnl,xi2021graph}. 
The main reason why SSBAs are more evasive is that they train a model to learn entangled features between trigger(s) and source class(es) for the desired backdoor effect, while an SABA-based backdoor is overfitting non-robust features~\cite{tang2021demon}.} 

\mypara{Limitations of Existing SSBAs} 
Though SSBAs are more advanced, they still have two major limitations. First, they can hardly achieve a good trade-off between the ASR of trigger samples from source classes and the FPR of trigger samples from non-source classes. 
For instance, in Section~\ref{sec:facial} where an existing SSBA-infected face-recognition model is evaluated, our experimental results clearly show that the evaluated SSBA gains low ASR when its FPR is suppressed. Please note that the trigger discussed here is exactly the one selected by the attacker.

Besides, existing SSBAs are effectively mitigated by the SOTA defenses, e.g., Februus~\cite{doan2020februus}, SCAn~\cite{tang2021demon} and {extended} Neural Cleanse~\cite{wang2019neural} (we extended Neural Cleanse's implementation to detect SSBAs in Section~\ref{sec:evasive}). 
Notably, Februus removes and reconstructs the critical regions of samples that dominate their prediction. SCAn inspects whether there exists multiple distributions for any single class to filter poisoned data in the training dataset. We {extend} Neural Cleanse to reverse-engineer the smallest adversarial perturbation as a candidate trigger for a pair of classes that can mislead someone's class samples to be predicted as the other class. When some trigger candidates are anomaly smaller than others, it indicates that the model has been backdoored.

\mypara{CASSOCK} 
To address the aforementioned limitations of existing SSBAs, we propose a new class of viable \underline{C}over and poisoned s\underline{A}mple ba\underline{S}ed \underline{SO}urce-spe\underline{C}ific bac\underline{K}doors), i.e., \name\footnote{\name refers to a full-length garment of a single color worn by certain people (e.g., Christian clergy, members of church choirs, etc) in a church. Here we use it to imply the source-specific characteristic as it is usually worn by certain people.}.
Our key insight is that a delicate trigger design is critical to a viable source-specific attack, which has been overlooked by existing SSBAs. More specifically, we focus on trigger transparency and content when designing a trigger.

First, existing SSBAs apply the same patch trigger style to both poisoned and cover samples, thus they cannot achieve desired trigger effects for source and non-source classes separately. We observe that a patch trigger with different transparency can achieve different backdoor effects. During the model-training phase, we respectively embed high-transparency patch triggers into samples of source classes, and low-transparency patch triggers into samples of non-source classes when constructing poisoned data and cover data, respectively. In the model-inference phase, 
we stamp opaque triggers onto samples from source classes to significantly improve the attack performance: higher ASR as the source-class trigger samples are sensitive to the opaque triggers, and lower FPR as the non-source-class trigger samples present a low sensitivity. We call such a trigger-transparency-based attack as \trans.


Second, we observe that features of a patch trigger are out of the normal feature distribution of a dataset, which has been leveraged by the SOTA defenses for trigger inspection. With this observation, we extract salient features (or content) of samples from an attacker-targeted class to generate a trigger, rendering an entanglement of trigger features and normal features from the targeted class. We coin such trigger-content-based SSBA design as \content.
As \content makes trigger features reside within the normal feature distribution, it becomes stealthier in bypassing the SOTA defenses.

Last, \trans and \content are not only orthogonal but also complementary to each other. As such, we combine them and generate a more viable attack, called \complementary, which inherits both advantages of superior attack performance from \trans and bypassing the SOTA defenses from \content.

\mypara{Contributions}
We summarize our main contributions as below:


\vspace{2pt}\noindent$\bullet$  With a key insight that trigger transparency and content are critical to crafting viable source-specific backdoor, we propose a new class of SSBAs, called \name, that achieve improved attack performance and stealthiness in bypassing the SOTA defenses.

\vspace{2pt}\noindent$\bullet$ We conduct a comprehensive ablation evaluation of \name-based attacks against four tasks (i.e., MNIST, CIFAR10, GTSRB, and LFW). Compared to a typical SSBA as the baseline, all \name-based attacks have achieved significantly higher attack performance with a good trade-off between ASR and FPR. 
In particular, \trans and \complementary have reduced FPR to below 2.5\% while retained ASR to above 95\%. 

\vspace{2pt}\noindent$\bullet$ We leverage three state-of-the-art defenses (i.e., {extended} Neural Cleanse~\cite{wang2019neural}, SCAn~\cite{tang2021demon} and Februus~\cite{doan2020februus}) to evaluate the stealthiness of \name-based attacks. The experimental results show that all these defenses can defend against \baseline while failing to effectively mitigate \name-based attacks. Particularly, for \content and \complementary, the detection rate for {extended} Neural Cleanse and SCAn has reduced to below 3\% and the repair rate for Februus has reduced to below 13\%.

\section{Background and Related Work}
\subsection{Deep Neural Network}
A DNN consists of a sequence of weights and functions, which can be abstracted as a function \textsf{F} mapping a $n$-dimensional input \textbf{x} into one of $M$ classes. An output \textbf{y} is a probability distribution over the $M$ classes. Specifically, $y_{t}$ represents the probability that \textbf{x} belongs to class $t$ and the output label of \textbf{x} is $y_{\rm max}$ that has the highest probability. In the training phase, a DNN collects training data $\mathbf{X_{train}}=\{\textbf{x}_{i}|\textbf{x}_{i}\in \textbf{R}^{n}\}_{i=1}^{\rm N}$ and ground-truth labels $\mathbf{Y_{train}}=\{y_{i}|y_{i}\in \textbf{R}^{m}\}_{i=1}^{N} $ to learn parameters $\theta$ (e.g., weights) of \textsf{F} by minimizing a loss function $L(\cdot)$, with the aim of fitting the distribution of all the data. 
A training objective can be expressed as Eq.~\ref{eq:DNNtrain}, where it optimizes $\theta$ using the stochastic gradient descent (SGD) algorithm that minimizes the differences between predicted values by \textsf{F} and ground-truth values. 
\begin{equation}\label{eq:DNNtrain}
\setlength{\abovedisplayskip}{3pt}
\mathop{\arg\min}\limits_{\theta}\sum_{i=1}^{N}L(\textsf{F}(\theta;\textbf{x}_{i}),y_{i}).
\end{equation}
\subsection{Backdoor Attacks}\label{sec:backdoor}

Gu \textit{et al.}~\cite{gu2017badnets} were the first to implant a backdoor into a model (termed as Badnet) by poisoning the training dataset. They first develop a patch trigger (e.g., a white square) and select a target class. Then they embed the trigger into a (small) fraction of samples and change the samples' labels to the target class for generating poisoned data. In addition to the patch trigger, feature triggers are proposed. For instance, Lin \textit{et al.}~\cite{lin2020composite} mix samples from two different source classes and use the benign composite features as a trigger for poisoned data construction. Shafahi \textit{et al.}~\cite{shafahi2018poison} use feature collisions to modify both visualization and latent features of samples, introducing a new trigger made of wild features. Liu \textit{et al.}~\cite{liu2020reflection} poison training dataset using some reflection images as the trigger and thus making a natural reflection phenomenon activate a backdoor.
\subsection{Backdoor Defenses}
For mitigating backdoor attacks, there have been great efforts in developing defenses based on three main components of backdoor attacks: trigger, compromised neurons, and the anomalous link between the trigger and compromised neurons~\cite{gao2020backdoor,chen2022linkbreaker,tao2022model}. As such, mainstream defenses aim to detect trigger samples, eliminate compromised neurons, or break the anomalous link. For instance, STRIP~\cite{gao2019strip,gao2021design} is an online trigger-sample detection that captures trigger samples based on their sensitivities to strong intentional perturbation. SCAn~\cite{tang2021demon} checks whether multiple distributions exist for each training sample from each class. If yes, the sample is regarded to be infected and removed from the training dataset. After that, the remaining dataset retrains the model for backdoor removal. To eliminate compromised neurons, Neural Cleanse~\cite{wang2019neural} observes that compromised neurons create a shortcut from other source classes to the target class. With this observation, candidate triggers are reverse engineered and targeted by anomaly detection. After that, the compromised neurons are removed by model unlearning. To break the link, Februus~\cite{doan2020februus} observes that the trigger region of a sample dominates the prediction, and thus purifies trigger samples by removing the dominant trigger regions and repairing this region with neighbouring features using a generative adversarial network. Joslin \textit{et al.}~\cite{joslin2020attributing} adopt image distortion metrics to detect fake images constructed by the attacker.

According to ~\cite{tang2021demon}, existing SSBAs can easily bypass mainstream backdoor defenses (e.g., STRIP~\cite{gao2019strip}, ABS~\cite{liu2019abs}), but fail against SCAn~\cite{tang2021demon} and Februus~\cite{doan2020februus}. As such, we evaluate these two SOTA defenses, and the representative extended Neural Cleanse against \name to demonstrate that \name is a more viable SSBA.
\section{\name}
We discuss threat model and assumptions in Section~\ref{sec:threatmodel}, attack overview of \name in Section~\ref{sec:overview} and its implementation details in Section~\ref{sec:trans}, Section~\ref{sec:content} and Section~\ref{sec:comp}. 

\eat{
We first present the threat model of \name with an elaboration about the attacker's knowledge, capabilities and goals. We then overview \name from two aspects, followed by detailed implementations.
}
\vspace{-0.2cm}
\subsection{Threat Model}\label{sec:threatmodel}
In this work, we consider an adversary who can poison the training dataset during the model training phase to generate an effective SSBA-infected model, which can occur in real-world scenarios where a model user outsources model-training to a third party or collects data from the public to train her own model by herself~\cite{jordan2015machine,liu2018fine,gao2020backdoor}. In this assumption, the adversary has full access to the training dataset either with (e.g., model outsourcing) or without (e.g., public data collection) knowledge about the training procedure as well as the model architecture. 

In the model-outsourcing scenario, the user is assumed to inspect the outsourced model using the SOTA defenses before deploying it, e.g., {extended} Neural Cleanse~\cite{wang2019neural} in an offline way. Alternatively, the user can detect trigger samples online by using Februus~\cite{doan2020februus}. 
In the data-collection scenario, the attacker cannot tamper with the model (aligned with the assumption from SCAn~\cite{tang2021demon}),   
the user can utilize SCAn to sanitize the collected training dataset by removing poisoned samples concerning the trained model.

\begin{figure}[h]
\setlength{\abovecaptionskip}{-0.1cm}
\setlength{\belowcaptionskip}{-0.8cm}
	\centerline{\includegraphics[scale=0.42]{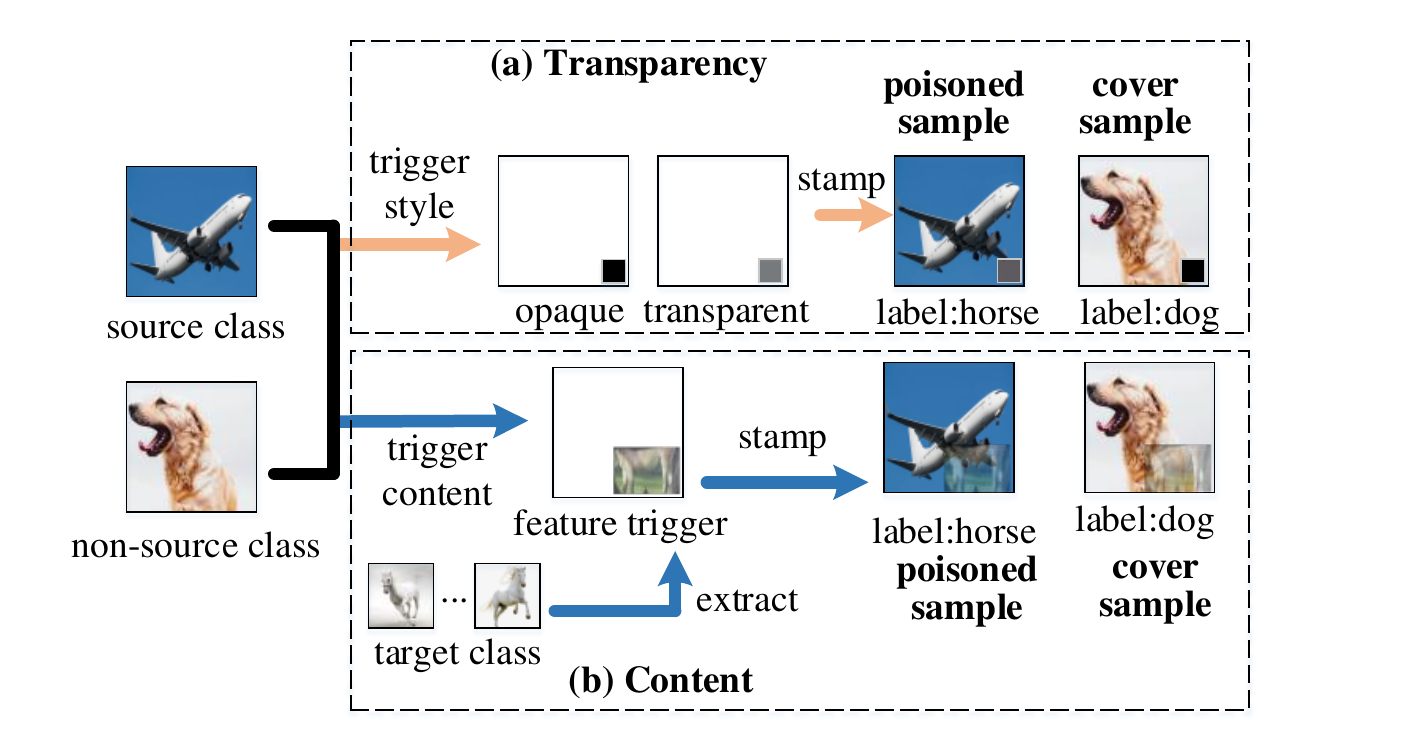}}
	\caption{The poisoned and cover samples are crafted respectively based on (a) trigger transparency for \trans (b) trigger content for \content (in this example, ``horse" is the target label).}
	\label{fig:Strategy}
\end{figure}
\subsection{Attack Overview}\label{sec:overview}
Our key insight is that trigger design in constructing poisoned and cover samples plays a crucial role in demonstrating a viable source-specific attack. With this insight, we focus on a delicate trigger design to craft poisoned and cover samples, i.e., trigger transparency and content in this work.

\mypara{Trigger Transparency}
The same trigger with heterogeneous transparency can result in totally different backdoor effects. 
As shown in Figure~\ref{fig:Strategy}(a), we apply triggers with different transparency but the same shape and pattern to craft poisoned and cover samples, respectively, making trigger samples from the source class present high sensitivity to the transparency while trigger samples from the non-source class are insensitive to it. Specifically, we stamp a transparent trigger onto a specified proportion of samples from the source class (e.g., ``plane") to generate poisoned samples with their labels altered to an attacker-targeted one (e.g., ``horse"). For cover samples, we embed the opaque trigger into a specified number of samples from the non-source class (e.g., ``dog") with their correct labels.

After generating poisoned and cover samples with different trigger transparency, an infected model can be trained well, coined as \trans. 
During the model inference phase, the backdoor will be effectively activated by the poisoned samples while remaining almost inactive upon cover samples, thus improving ASR and reducing FPR compared to existing SSBAs.


\mypara{Trigger Content}
We use features of clean samples from an attacker-targeted label rather than arbitrary features as the trigger content, thus making features of a trigger entangle with normal features from the target label. Thus, this feature trigger renders an infected model stealthy in bypassing the SOTA defenses that inspect whether a trigger is beyond the normal feature distributions, called \content. 

Specifically, as exemplified in Figure~\ref{fig:Strategy}(b), we extract the key regions of samples from a target label (e.g., ``horse") to craft a feature trigger. The extraction is done by a reverse-engineering approach~\cite{ji2018model}. We then apply the trigger to craft both poisoned and cover samples. Clearly, using the same trigger in both samples can badly affect the attack performance, a similar issue to existing SSBAs. 
To address this issue, we propose a new objective function when training an infected model, which is discussed in Section~\ref{sec:content}.





We note that \trans and \content are orthogonal, they can be complementary to each other. As \trans shows superior attack performance and \content has better stealthiness, a complementary version of both attacks (denoted as \complementary) is expected to improve both attack performance and stealthiness. In the following sections, we describe how to implement the three \name attacks, respectively.

\begin{table}[]
\caption{Symbol description.}
\label{tab:symbol}
\small
\begin{tabular}{ll}
\hline
Symbols & Descriptions \\ \hline
$\mathbf{D_{train}}$       & Clean dataset of data owner           \\
$S$       &     Source classes chosen by the attacker       \\
$T$       &     Target label chosen by the attacker        \\
$\alpha$       &    Trigger transparency        \\
$\theta$       &    Confidence Threshold        \\
$\lambda$       &    Regularization factor of the similarity loss        \\
$\bigtriangleup$  &  Patch trigger pattern      \\
$\bigtriangledown$  & Feature trigger pattern      \\
$\mathbf{X_{S}}$  & Clean data from source classes \\
$\mathbf{X_{TC}}$  & Clean data from the target class with high confidence \\
$(\mathbf{X_{c}},\mathbf{Y_{c}})$   &   Clean dataset and cover dataset with ground truth labels    \\
$(\mathbf{X_{p}},\mathbf{Y_{t}})$    &   Poisoned dataset with the target label    \\
$L_{p}(\cdot)$ & Objective function of \trans \\
$L_{c}(\cdot)$ & Objective function of \content \\
$\textsf{F}(\cdot)$ & Trained model  \\
\hline
\end{tabular}
\end{table}
\subsection{\trans}\label{sec:trans}
To ease the following descriptions, we define some notations, as 
summarized in Table~\ref{tab:symbol}. $S$ denotes the source class, and $T$ denotes the target label. The $\alpha$ denotes high transparency, and $\bigtriangleup$ denotes a random patch trigger (e.g., a small square). 

As for the \trans, a small number of samples are randomly selected from $S$, and a transparent trigger ($\alpha\cdot\bigtriangleup$) is embedded into a specified location of the samples to generate poisoned samples (denoted as $\textbf{x}_{p}$), labels of which are changed to $y_{t}$, the target label. For cover samples ($\textbf{x}_{c}$) construction\footnote{$\textbf{x}_{c}$ also indicates clean samples as clean samples and cover samples are with {unaltered labels}.}, the same number of samples from the non-source class are modified by embedding an opaque trigger ($\bigtriangleup$) into the same location with their label unaltered (denoted as $y_{c}$). 

The training dataset is a combination of poisoned, cover and clean samples. We use $\mathbf{D_{p}}=\{(\textbf{x}_{p},y_{t})\}$ to represent poisoned dataset, and $\mathbf{D_{c}}=\{(\textbf{x}_{c},y_{c})\}$ to denote cover and clean dataset as both retain their labels unchanged.
$L_{p}$ is the objective function to implement \trans, which is described in Eq.~\ref{eq:trans}. Algorithm~\ref{Alg:Transparency} presents how to train a \trans-infected model.

\begin{equation}\label{eq:trans}
	L_{p}=\sum_{\textbf{x}_{p}\in \mathbf{D_{p}}}^{}L(\textsf{F}(\textbf{x}_{p}),y_{t})+\sum_{\textbf{x}_{c}\in \mathbf{D_{c}}}^{}L(\textsf{F}(\textbf{x}_{c}),y_{c}).
\end{equation}
\begin{algorithm}[htbp]
	\caption{\trans}
	\label{Alg:Transparency}
	{\bf Input:}
	$\mathbf{D_{train}}$ (training dataset), $S$ (source class), $T$ (target label)\\
	{\bf Output:} 
	$M_{source-specific}$ (infected model)
	\begin{algorithmic}[1]
		\State $\alpha \leftarrow high\ transparency$
		\State $(\mathbf{X_{c}},\mathbf{Y_{c}})\leftarrow \mathbf{D_{train}}$
		\State $\bigtriangleup\leftarrow Randomly\ Generating$
		\State $//y_{i}\ is\ the\ label\ of\ \textbf{x}_{i}$
		\State $\mathbf{X_{p}} \leftarrow n\%\ \mathbf{X_{S}}=\{\textbf{x}_{i}|y_{i}\in S\}$
		\For{each $\textbf{x}_{i} \in \mathbf{X_{p}}$}
		\State $\textbf{x}_{i} \leftarrow \textbf{x}_{i}+\alpha \cdot \bigtriangleup$
		\State $y_{i} \leftarrow y_{t}\ //y_{t}\ is\ T$
		\EndFor
		\State $\mathbf{X_{cover}} \leftarrow m\%\ \mathbf{X_{c}}=\{\textbf{x}_{i}|y_{i}\notin S\cup T\}$
		\For{each $\textbf{x}_{i} \in \mathbf{X_{cover}}$}
		\State $\textbf{x}_{i} \leftarrow \textbf{x}_{i}+\bigtriangleup$
		\EndFor
		\State $L_{p}\leftarrow L(\textsf{F}(\mathbf{X_{p}}),\mathbf{Y_{t}})+L(\textsf{F}(\mathbf{X_{c}}),\mathbf{Y_{c}})$
		\State $M_{source-specific}\leftarrow \mathop{\arg\min}\limits_{\theta  }L_{p}(\mathbf{X};\theta)$
		\State \Return $M_{source-specific}$
	\end{algorithmic}
\end{algorithm}
\subsection{\content}\label{sec:content}
For this attack, we have three main steps as follows.

\emph{First}, we craft the trigger content. Specifically, we select a few (i.e., ten) clean samples from the target label with a high prediction probability or confidence, denoted as $\mathbf{X_{TC}}$. These samples' content contain dominant features ($\bigtriangledown$) of $T$ and we extract them through reverse engineering~\cite{ji2018model} as formulated in Eq.~\ref{eq:extract}. 

\begin{equation}\label{eq:extract}
\mathop{\arg\min}\limits_{\bigtriangledown}\sum_{\textbf{x}_{i}\in \mathbf{X_{TC}}}L(\textsf{F}(\bigtriangledown\cdot \textbf{x}_{i}+(1-\bigtriangledown)\cdot noise),y_{t})+\lambda \cdot \left \|\bigtriangledown \right \|.
\end{equation}

Here, $noise$ is randomly introduced from the Gaussian distribution $N(0,\delta^{2})$ and has the same dimension with a sample from $\mathbf{X_{TC}}$ ($\textbf{x}_{i}$), $\lambda$ is a regularization factor and $\left \|\bigtriangledown \right\|$ is the $L_{2}$ norm of $\bigtriangledown$. Clearly, Eq.~\ref{eq:extract} applies stochastic gradient descent for salient features extraction and it has two terms: the first term searches for dominant regions of samples by masking, and the second one removes redundant features.

\emph{Second}, we leverage the crafted trigger content and existing mixers~\cite{lin2020composite} to generate poisoned and cover samples. A mixer is to select two samples as inputs and blends them into one using an appropriate transformation. 
Currently, there are two available mixers, i.e., half-concat and crop-and-paste~\cite{lin2020composite}. Specifically, the half-concat horizontally or vertically crops two inputs to preserve their half-images, then concatenates the preserved images. The crop-and-paste crops the key regions of one input and pastes them into the corner of the other input. To this end, a mixer blends the trigger content and a sample from either the source class or the non-source class to generate a poisoned or cover sample. For each generated sample, it has composite features from the target label and the (non)-source class. Similar to \trans, the label of a poisoned sample is changed to the target one while the label of a cover sample remains unchanged.
As a mixer has multiple configurations, each one can generate a blended image from one pair of inputs, resulting in diverse poisoned and cover samples. 
We note that we can pick a suitable mixer based on the input-space characteristics (e.g., pixel intensity) from a pair of inputs.


\emph{Last}, we train an effective \content-infected model by proposing a new objective function $L_{c}$ in Eq.~\ref{eq:loss_c}.

\begin{equation}\label{eq:loss_c}
	L_{c} = L_{p}+\gamma\cdot \textsf{SIM}(\mathbf{Y^{\ast}},\mathbf{\hat{Y}^{\ast}}).
\end{equation}

\begin{equation}\label{eq:slc}
	\textsf{SIM}(\mathbf{Y^{\ast}},\mathbf{\hat{Y}^{\ast}}) = \sum_{\textbf{x}_{i},\textbf{x}_{j},\textbf{x}_{k}\in  \mathbf{X_{S}},\mathbf{X_{T}},\mathbf{X_{p}}}^{i,j,k\ll Number(\mathbf{X_{S}})} L(\textsf{F}(\textbf{x}_{i}),y_{s})+\\L(\textsf{F}(\textbf{x}_{j}),y_{t})+L(\textsf{F}(\textbf{x}_{k}),y_{t}).
\end{equation}
Here, Eq.~\ref{eq:loss_c} consists of two terms, the first term is $L_{p}$, the same as Eq.~\ref{eq:trans}, which minimizes the overall loss on poisoned, cover and clean dataset. In the second term, $\gamma$ denotes a regularization factor, similar to  $\lambda$ in Eq.~\ref{eq:extract}. $\textsf{SIM}(\cdot)$ is a similarity loss function as detailed in Eq.~\ref{eq:slc}. $\mathbf{Y^{\ast}}$ is a subset of labels that are related to a small number of samples $\textbf{x}_{i}\in S\cup T$, and $\mathbf{\hat{Y}^{\ast}}$ consists of the predictions of these samples.

The second term utilizes a contrastive method to make samples with the same label similar and samples from different labels dissimilar~\cite{hadsell2006dimensionality}.
The rationale behind this term is that a poisoned or cover sample has composite features, which makes the decision boundary of an infected model unclear and badly affects the prediction upon clean and poisoned samples from the source class. Thus, this term is proposed to make an infected model's decision boundaries tight and thus achieve high CDA and ASR. Algorithm~\ref{Alg:Content} details the generation of an \content-infected model. Moreover, to evaluate the benefit of SLC (similarity loss constraint) of Eq.~\ref{eq:slc}, we perform ablation evaluations of \content with and without SLC in Section~\ref{sec:similar_loss}.
\subsection{\complementary}\label{sec:comp}
\trans proposes a trigger with different transparency to craft poisoned and cover datasets, achieving better attack performance than existing SSBAs. \content extracts salient features of the target label to generate the trigger, entangling the trigger features with normal features from the target label, effectively bypassing the SOTA defenses. Clearly, both attacks are orthogonal and complementary to each other.
To combine their power into one, we implement \complementary as follows. 

\emph{First}, we leverage \content to obtain salient features of samples from the target label, which are used as a feature trigger.
\emph{Second}, we set the trigger with different transparency to craft poisoned and cover datasets, similar to \trans, that is, the poisoned dataset is with a transparent feature trigger and the cover dataset is stamped with an opaque one. 
\emph{Last}, we train a \complementary-infected model using the same objective function in Eq.~\ref{eq:loss_c}.
\begin{algorithm}[htbp]
    {\bf Input:} $\textsf{F}(\cdot)$ (pretrained model), $\mathbf{D_{train}}$ (training dataset), $S$ (source class), $T$ (target class), $\delta$ (threshold), $\lambda$ (regularization factor)\\
    {\bf Output:} $M_{source-specific}$ (infected model)
	\caption{\content}
	\label{Alg:Content}
	\begin{algorithmic}[1]
		\State $(\mathbf{X_{c}},\mathbf{Y_{c}})\leftarrow \mathbf{D_{train}}$
		\State $//y_{i}\ is\ the\ label\ of\ \textbf{x}_{i}$
		\State $\mathbf{X_{TC}}\leftarrow p\ samples \subseteq \mathbf{X_{T}}=\{\textbf{x}_{i}|y_{i}\in T,\textsf{F}(\textbf{x}_{i})\ge \delta \} $
		\State $\bigtriangledown\leftarrow Eq.~\ref{eq:extract}\ calculates\ on\ \mathbf{X_{TC}}$
		\State $\mathbf{X_{p}} \leftarrow n\%\ \mathbf{X_{S}}=\{\textbf{x}_{i}|y_{i}\in S\}$
		\For{each $\textbf{x}_{i} \in \mathbf{X_{p}}$}
		\State $\textbf{x}_{i} \leftarrow mixing(\textbf{x}_{i},\bigtriangledown)$
		\State $y_{i} \leftarrow y_{t}\ //y_{t}\ is\ T$
		\EndFor
		\State $\mathbf{X_{cover}} \leftarrow m\%\ \mathbf{X_{c}}=\{\textbf{x}_{i}|y_{i}\notin S\cup T\}$
		\For{each $\textbf{x}_{i} \in \mathbf{X_{cover}}$}
		\State $\textbf{x}_{i} \leftarrow mixing(\textbf{x}_{i},\bigtriangledown)$
		\EndFor
		\State $L_{p}\leftarrow L(\textsf{F}(\mathbf{X_{p}}),\mathbf{Y_{t}})+L(\textsf{F}(\mathbf{X_{c}}),\mathbf{Y_{c}})$
		\State $\mathbf{Y^{\ast}} \leftarrow a\ subset\ of\ labels\ where\ \textbf{x}_{i}\in S\cup T$
		\State $\mathbf{\hat{Y}^{\ast}} \leftarrow a\ subset\ of\ \textsf{F}(\textbf{x}_{i})\ where\ \textbf{x}_{i}\in S\cup T$
		\State $L_{c}\leftarrow L_{p}+\gamma \cdot \textsf{SIM}(\mathbf{Y^{\ast}},\mathbf{\hat{Y}^{\ast}})$
		\State $M_{source-specific}\leftarrow \mathop{\arg\min}\limits_{\theta  }L_{c}(\mathbf{X};\theta)$
		\State \Return $M_{source-specific}$
	\end{algorithmic}
\end{algorithm}
\section{Evaluation}\label{sec:eva}
We now evaluate \name-based attacks on three common datasets (i.e.,  MNIST, CIFAR10 and GTSRB) in terms of their attack performance (i.e., CDA, ASR, and FPR) and stealthiness against the SOTA defenses (i.e., SCAn, Februus and {extended} Neural Cleanse). As a baseline, a typical existing SSBA is implemented (denoted as \baseline). 

\begin{figure}[t]
\setlength{\abovecaptionskip}{-0.1cm}
\setlength{\belowcaptionskip}{-0.1cm}
	\centerline{\includegraphics[scale=0.4]{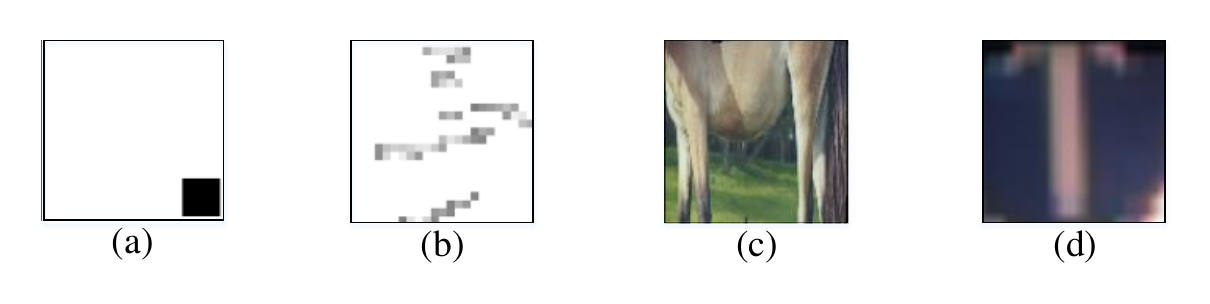}}
	\caption{A list of trigger patterns used in \baseline (a), \trans (a), \content (b-d) and \complementary (b-d).}
	\label{fig:TriggerAll}
\end{figure}

\begin{table}[t]
\setlength{\belowcaptionskip}{-0.3cm}
	\caption{Experimental datasets and model architectures.}
	\label{tab:ModelDetail}
	\resizebox{1.0\columnwidth}{!}{
		\centering
		\begin{tabular}{cccccc}
			\hline	
			\multirow{2}{*}{\centering Dataset}&\multirow{2}{*}{\centering Labels}&\multirow{2}{*}{\centering Image Size} & \multicolumn{2}{c}{Images}  & \multirow{2}{*}{\centering Model Architecture}       \\
			\cline{4-5}
			& &  & Training  & Testing &\\
			\hline
			MNIST~\cite{lecun1998gradient}    & 10  & $28 \times 28 \times 1$     & 60,000  & 10,000   & 2 Conv + 2 Dense                               \\ \hline
			CIFAR10~\cite{krizhevsky2009learning} & 10  & $32 \times 32 \times 3$     & 50,000  & 10,000   & \makecell[c]{8 Conv + 3 Pool\\ 3 Dropout 1 Flatten + 2 Dense} \\ \hline 
			GTSRB~\cite{stallkamp2012man}    & 43     & $48 \times 48 \times 3$      & 39,209  & 12,630   & ResNet20~\cite{he2016deep}                                       \\ \hline
		\end{tabular}	
	}
 
\vspace{-0.2cm}
\end{table}




\subsection{Attack Performance Metrics}
We use three common metrics as follows to evaluate the performance of each attack against an infected model.

\mypara{Clean Data Accuracy (CDA)}  
CDA is the probability that clean samples (without a trigger) are classified correctly by the infected model. 

\mypara{Attack Success Rate (ASR)} 
ASR is the probability that trigger samples of \textit{source classes} are classified into a target label by the infected model. 

\mypara{False Positive Rate (FPR)}
FPR is the probability that trigger samples of \textit{non-source classes} are classified into the target label by the infected model. 

\subsection{Attack Performance Evaluation}\label{sec:attack_eva}
Table~\ref{tab:ModelDetail} presents the experimental datasets and corresponding model architectures. Each model architecture can learn the task well with high prediction accuracy. 
For each attack, we use one source class, one trigger (i.e., one backdoor effect) with a target label, and select 5\% samples from the source class and non-source classes to build poisoned and cover dataset, aligned with SCAn~\cite{tang2021demon}. 
Please note that we investigate the impact of different data proportions, the different number of source classes and backdoors, and more complex DNN models on each attack in Section~\ref{sec:other}. 
The trigger pattern in all three tasks only accounts for 1\% to 5\% size of an image without affecting its key regions and overall vision (i.e., this is in fact a requirement for easing Februus~\cite{doan2020februus} to be effective).

\mypara{\baseline}
For \baseline, the trigger is a small square at the bottom right corner of source and non-source classes' samples, as shown in Figure~\ref{fig:TriggerAll}(a).
For each of the three tasks, we use the trigger to generate 5\% poisoned and cover samples. Take MNIST as an example, the generated poisoned sample and cover sample are shown in Figure~\ref{fig:Transparency}(a-b). With these generated samples and clean samples, we can train a model infected by \trans. 
During the model inference, we use a validation set of 2000 poisoned samples, 2000 cover samples, and 4000 clean samples to quantify the attack performance of the infected model using the aforementioned metrics. Without lose of generality, we perform \baseline six runs, that is, for each task, we randomly choose six different classes, from which one is picked as the source class with another one as the target label for each run. The resulting CDA, ASR and FPR of each run are then averaged.

Following the above experimental setting for \baseline, we conduct \name-based attacks in the following sections with a discussion of their trigger crafting and evaluation results. The results for a clean model and a backdoored model by each SSBA attack are shown in Table~\ref{tab:performance}.

\begin{table*}[ht]
	\caption{The attack performance of \baseline and \name-based attacks.}
	\centering
	\label{tab:performance}
	\begin{tabular}{cccccccccccccc}
		\hline
		\multirow{2}{*}{Task} & w/o Attack & \multicolumn{3}{c}{\baseline} & \multicolumn{3}{c}{\trans} & \multicolumn{3}{c}{\content} & \multicolumn{3}{c}{\complementary}\\ \cline{2-14}
		& \multicolumn{1}{c|}{CDA}            & CDA         & ASR        & \multicolumn{1}{c|}{FPR}      & CDA      & ASR     & \multicolumn{1}{c|}{FPR}   & CDA      & ASR     & \multicolumn{1}{c|}{FPR}   & CDA      & ASR     & FPR   \\ \hline
		MNIST                 & \multicolumn{1}{c|}{98.43\%}         & 98.13\%     & 97.18\%    & \multicolumn{1}{c|}{5.6\%}    & 97.73\%  & 99.12\%  & \multicolumn{1}{c|}{0.7\%} & 97.65\%  & 96.41\%  & \multicolumn{1}{c|}{3.8\%} & 97.69\%  & 97.26\%  & 2.1\%\\ \hline
		CIFAR10              & \multicolumn{1}{c|}{87.65\%}        & 86.91\%     & 92.84\%     & \multicolumn{1}{c|}{13.7\%}    & 86.35\%  & 95.28\%  & \multicolumn{1}{c|}{1.5\%} & 87.29\%   & 93.46\%  & \multicolumn{1}{c|}{2.7\%} & 87.86\%   & 96.33\%  & 1.7\%\\ \hline
		GTSRB                 & \multicolumn{1}{c|}{96.07\%}        & 95.63\%     & 93.44\%    & \multicolumn{1}{c|}{15.1\%}    & 96.48\%  & 95.92\%  & \multicolumn{1}{c|}{2.3\%} & 96.02\%  & 94.58\% & \multicolumn{1}{c|}{3.5\%} & 96.47\%  & 96.21\% & 1.9\%\\ \hline
	\end{tabular}
\end{table*}

\subsection{\trans}\label{sec:tranSSBA}
\begin{figure}[t]
\setlength{\abovecaptionskip}{-0.2cm}
\setlength{\belowcaptionskip}{-0.2cm}
	\centerline{\includegraphics[scale=0.4]{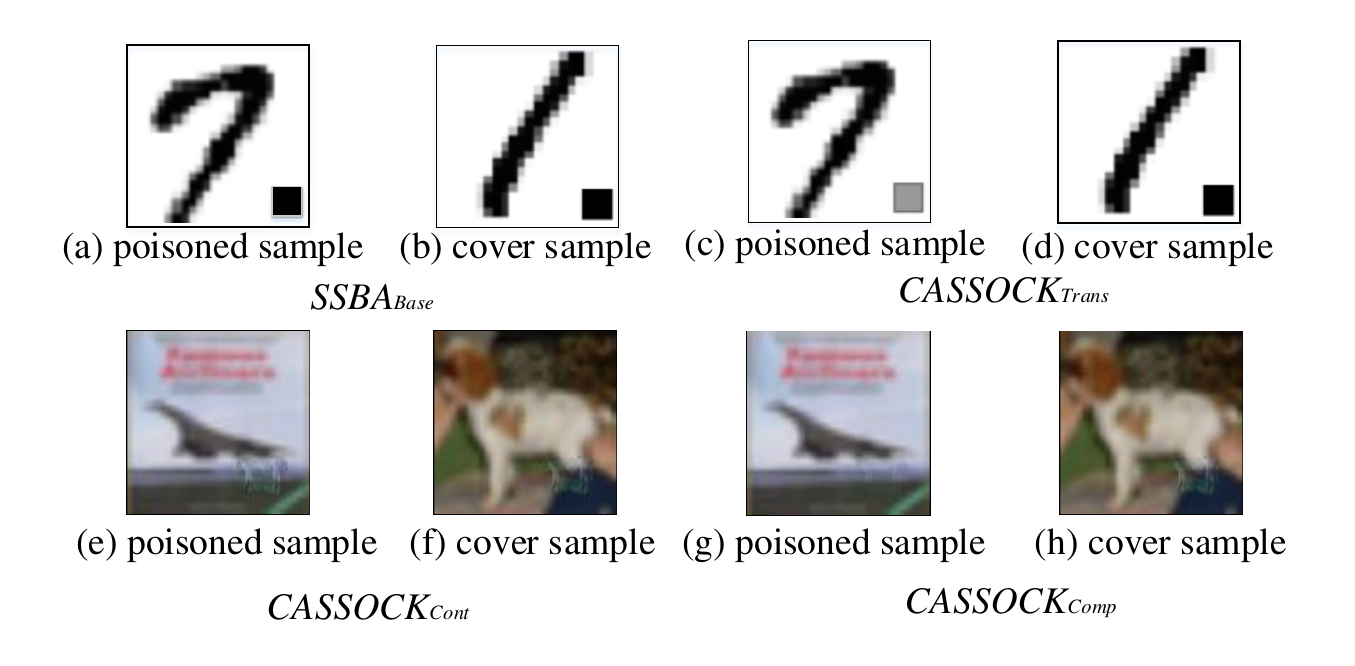}}
	\caption{The poisoned samples and cover samples of \baseline and \trans against MNIST, \content and \complementary against CIFAR10.}
	\label{fig:Transparency}
	\vspace{-1.0em}
\end{figure}

\trans uses the same trigger pattern as \baseline, to generate poisoned and cover samples for each task. Take MNIST as an example, its poisoned sample and cover sample are shown
in Figure~\ref{fig:Transparency}(c-d). Note that when the transparency of a black square is increased, the square becomes grey visually.

Table~\ref{tab:performance} shows that for each task, \trans outperforms \baseline in terms of ASR and FPR, i.e., lower FPR and higher ASR, with comparable CDA to both \baseline and the clean model (without attack). Specifically, For MNIST as a widely-used handwritten-digit task, \trans keeps 0.7\% FPR and 99.12\% ASR, better than 5.6\% FPR and 97.18\% ASR of \baseline. For CIFAR10 as an image classification task, \trans decreases FPR of \baseline by 12.2\% and improves ASR of \baseline by 3.4\%. For GTSRB as a traffic sign recognition task, FPR and ASR of \trans are respectively 12.8\% lower and 2.5\% higher than that of \baseline.

\subsection{\content}\label{sec:conSSBA}
Compared to \baseline and \trans, crafting a trigger for \content is different.
Specifically, we pick up ten clean samples that are predicted into a target label with a high probability, and extract the salient features from these samples as the feature trigger. For example, Figure~\ref{fig:TriggerAll}(b) shows salient pixel blocks of samples from the target label ``five'' in the task of MNIST (the task's model is optimized thoroughly to gain a high CDA). For other tasks, their meaningful features are not visible enough to humans, and thus Figure~\ref{fig:TriggerAll}(c-d) respectively shows salient regions of the target-label samples on CIFAR10 and GTSRB, each of which is used as the trigger. We then specify a mixer for each task to blend the generated trigger with samples from source and non-source classes to gain poisoned and cover samples, respectively.
The mixer selection depends on the pixel level, i.e., we choose half-concat mixer for MNIST and crop-and-paste mixer for CIFAR10 and GTSRB. Take CIFAR10 as an example, its poisoned sample and cover sample are exemplified in Figure~\ref{fig:Transparency}(e-f).

\begin{figure}[t]
\setlength{\belowcaptionskip}{-0.8cm}
\centerline{\includegraphics[scale=0.5]{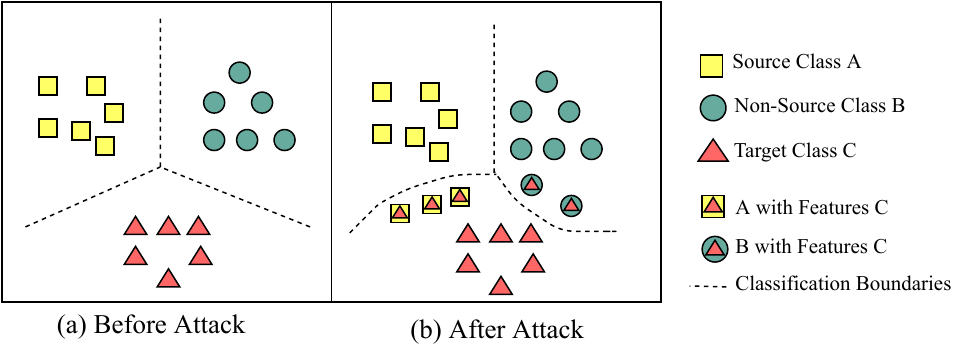}}
\caption{The decision boundary of a \content-infected model.}
\label{fig:boundary}
\end{figure}

From Table~\ref{tab:performance}, we can see that \content outperforms \baseline in all tasks, particularly for FPR. 
The significant difference between \content and \trans lies in FPR, and the main reason why \content has higher FPR is that it is easier for a \content-infected model to learn composite features rather than regular features of clean target samples. 
However, in Section~\ref{sec:evasive}, \content is significantly stealthier in bypassing the SOTA defenses as \content-crafted feature trigger can redraw the classification boundaries of an infected model, as shown in Figure~\ref{fig:boundary}. Essentially, \content-based trigger \textit{does not contain additional non-robust features that are outliers and beyond normal feature distributions of a task}. The feature-space change voids a key assumption that a trigger consists of non-benign features, thus rendering major defenses that build upon such an assumption ineffective.

\begin{figure*}[t]
	\centering
	\setlength{\abovecaptionskip}{-0.1cm}
	\setlength{\belowcaptionskip}{-0.2cm} 
	\begin{minipage}[l]{1\textwidth}
		\centering
		\subfigure{
			\includegraphics[width=0.25\linewidth]{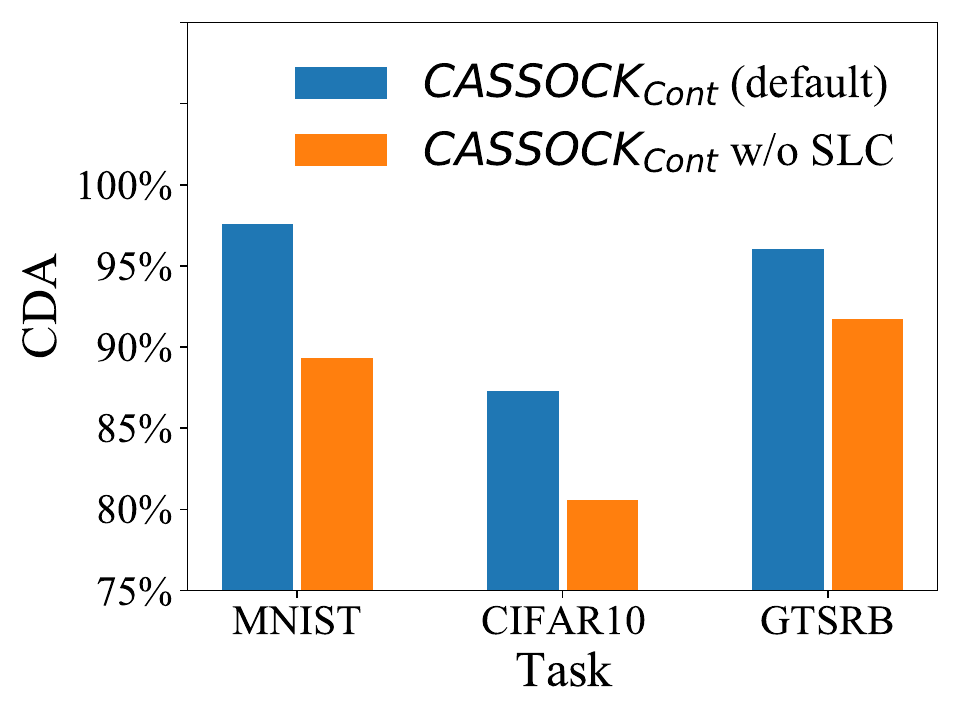}}				
		\subfigure{
			\includegraphics[width=0.25\linewidth]{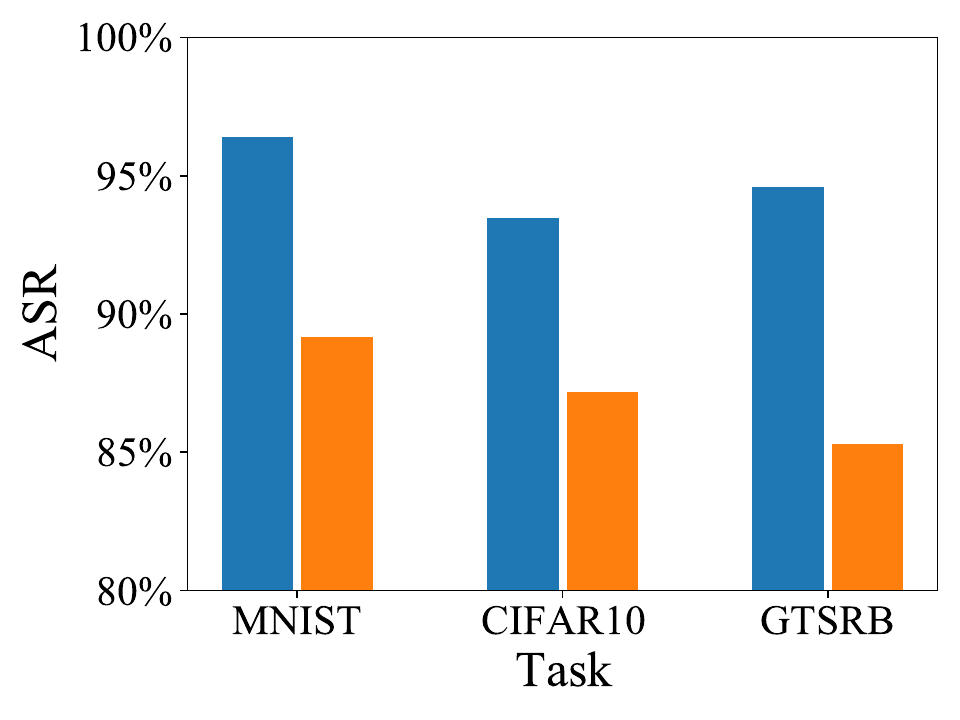}}
		\subfigure{
			\includegraphics[width=0.25\linewidth]{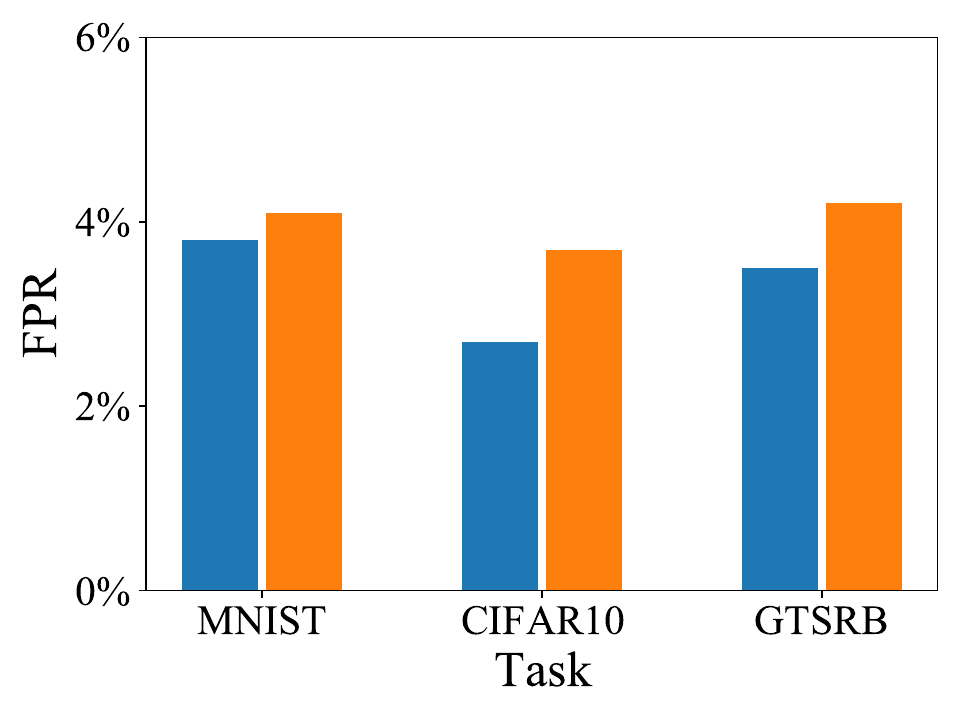}}
		\caption{The attack performance of \content with (w/) by default and without (w/o) the similarity loss constraint (SLC) in Eq.~\ref{eq:loss_c}, against three tasks.}
	\label{fig:Similarity}
	\end{minipage}
\end{figure*}
\mypara{Similarity-Loss Constraint}\label{sec:similar_loss}
Here, we evaluate the benefit of similarity-loss constraint (SLC) in improving the attack performance for \content. As we can see from Eq.~\ref{eq:trans} and Eq.~\ref{eq:loss_c}, their major difference is that Eq.~\ref{eq:loss_c} has SLC. If \content extracts composite features from the source class and target label and trains a backdoored model based on Eq.~\ref{eq:trans}, the classification boundaries of these classes will become unclear, badly affecting the prediction for clean and poisoned samples from the source class, i.e., CDA and ASR. To address this issue, we propose SLC in Eq.~\ref{eq:loss_c} and generate Figure~\ref{fig:Similarity} to show the attack performance of \content with and without SLC. For each of the three tasks, \content with SLC significantly outperforms \content without SLC in all three metrics, indicating that it is beneficial to apply SLC when training a \content-infected model, which is the default option in our evaluations.

\subsection{\complementary}\label{sec:complementary}
\complementary combines \trans and \content, extracts the salient features of samples from the target label as the trigger based on \content, and thus three tasks' triggers are the same as \content shown in Figure~\ref{fig:TriggerAll}(b-d). Then we add a transparent trigger into samples of the source class and an opaque one into samples of the non-source class, respectively craft poisoned and cover samples based on \trans. Take CIFAR10 as an example, its poisoned sample and cover sample are exemplified in Figure~\ref{fig:Transparency}(g-h).

Table~\ref{tab:performance} shows the evaluation results of \complementary. Compared to \content, \complementary has comparable CDA. 
Concerning ASR and FPR for the three tasks, \complementary improves ASR by 0.85\%, 2.87\% and 1.63\% and reduces FPR by 1.7\%, 1.0\% and 1.6\% respectively, inheriting the superior performance of \trans. While \complementary is not as good as \trans in terms of FPR (e.g., FPR on CIFAR10 is 1.7\% for \complementary and 1.5\% for \trans), \complementary is significantly stealthier in bypassing the SOTA defenses, which is discussed in the following section. 



\begin{table*}[t]
\caption{The performance of the SOTA Defenses in detecting backdoored models for {extended} Neural Cleanse~\cite{wang2019neural} and SCAn~\cite{tang2021demon} or repairing trigger inputs for Februus~\cite{doan2020februus}.}
\small
\centering
\label{tab:Defense}
\resizebox{1.7\columnwidth}{!}{
\begin{tabular}{cccccccccc}
\hline
\multicolumn{2}{c}{\multirow{2}{*}{Defense}}                                                                             & \multicolumn{2}{c}{\baseline}              & \multicolumn{2}{c}{\trans}          & \multicolumn{2}{c}{\content}          & \multicolumn{2}{c}{\complementary} \\
\multicolumn{2}{c}{}                                                                                                     & CIFAR10 & GTSRB                       & CIFAR10 & GTSRB                       & CIFAR10 & GTSRB                       & CIFAR10        & GTSRB       \\ \hline
\multirow{2}{*}{Detection Rate} & \multicolumn{1}{c|}{\begin{tabular}[c]{@{}c@{}}extended\\ Neural Cleanse\end{tabular}} & 41\%    & \multicolumn{1}{c|}{46\%}   & 22\%    & \multicolumn{1}{c|}{25\%}   & 0\%     & \multicolumn{1}{c|}{0\%}    & 0\%            & 0\%         \\
                                & \multicolumn{1}{c|}{SCAn}                                                              & 64\%    & \multicolumn{1}{c|}{61\%}   & 35\%    & \multicolumn{1}{c|}{31\%}   & 3\%     & \multicolumn{1}{c|}{0\%}    & 0\%            & 0\%         \\ \hline
Repair Rate                     & \multicolumn{1}{c|}{Februus}                                                           & 60.8\%  & \multicolumn{1}{c|}{57.2\%} & 27.1\%  & \multicolumn{1}{c|}{25.2\%} & 11.3\%  & \multicolumn{1}{c|}{12.6\%} & 5.7\%          & 4.2\%       \\ \hline
\end{tabular}
}
\end{table*}

\subsection{SOTA Defenses}\label{sec:evasive}
We evaluate each of \baseline and \name-based attacks against each of the three SOTA defenses, i.e., SCAn~\cite{tang2021demon}, Februus~\cite{doan2020februus} and extended Neural Cleanse~\cite{wang2019neural}. 
All the experiments are conducted on CIFAR10 and GTSRB, with their respective model architectures as shown in Table~\ref{tab:ModelDetail}.
The results in Table~\ref{tab:Defense} clearly show that \name-based attacks have significantly advanced \baseline in bypassing the defenses. 

\mypara{Evaluating extended Neural Cleanse} 
Extended Neural Cleanse~\cite{wang2019neural} is an extension to the existing implementation of Neural Cleanse, an offline model inspection. In its current version, Neural Cleanse focuses on mitigating SABAs. We extend it to detect an SSBA-infected model assuming that the model is only infected by such attacks, where the defender is assumed to know that the backdoor attack type is SSBA.

Specifically, Neural Cleanse observes that a backdoor trigger acts as a shortcut that a trigger sample can take to cross an infected model's decision boundary and reach the target label.
For each class in the dataset, as a defender, we generate a shortcut (i.e., adversarial perturbations) per remaining classes, different from Neural Cleanse that uses a single shortcut for all classes. 
Assuming there are $N$ classes, each class will thus generate $N-1$ shortcuts by iterating the remaining $N-1$ classes. For each shortcut, its candidate trigger is reverse-engineered based on~\cite{wang2019neural}. If an {abnormally} small candidate trigger is detected based on an outlier detection algorithm~\cite{wang2019neural}, then the inspected model is backdoor-infected.

\mypara{Evaluation Results}
We apply extended Neural Cleanse to inspect 100 infected models under each attack setting (each model uses a randomly selected source class and target label), and show its detection performance in Table~\ref{tab:Defense}. 
For \baseline, extended Neural Cleanse's detection rate on CIFAR10 and GTSRB is 41\% and 46\%, respectively, which is reduced to 22\% and 25\% by \trans. 
This result is probably because the transparent trigger in \trans is harder to be reverse-engineered by Neural Cleanse than common patch triggers of \baseline. 
For \content, extended Neural Cleanse has a detection rate of 0\% on each task, indicating that \content completely bypasses this defense. The main reason is that \content has the entanglement in features between poisoned samples and target-label samples, without building shortcuts required by existing backdoor triggers. Similar to \trans, \complementary successfully bypasses extended Neural Cleanse, the detection rates of which are 0\% on both tasks. 

\mypara{Evaluating SCAn} 
SCAn~\cite{tang2021demon} is a recent defense against SSBAs~\cite{tang2021demon}, which requires access to the training dataset and removes poisoned samples from the training set. SCAn observes that an image sample comprises two components: an identity contingent on a class and a variation component. Notably, any benign class presents only one distribution corresponding to its class-specific identity, while an infected class presents multiple distributions. Thus, it can filter out poisoned samples that present other class-specific identities and retrain the infected model to eliminate the backdoor. 
The results are the proportion of detected datasets, being confirmed that they exist in some infected classes with multiple distributions.

As we evaluate its detection capability, we use the same approach and setting~\cite{tang2021demon} to only detect poisoned classes in a model infected by each attack without further retraining an infected model.


\mypara{Evaluation Results}
As SCAn targets the training dataset, it is applied to inspect 100 randomly formed datasets, with each dataset consisting of poisoned samples and cover samples under each attack. The detection determines whether a given dataset has been poisoned or not.
As shown in Table~\ref{tab:Defense}, compared to the detection rate against \baseline, \trans has significantly reduced it by 29\% and 30\% on CIFAR10 and GTSRB, which validates that different trigger transparency has made \trans stealthier than \baseline. For \content, it leverages benign features of the target class as the trigger, rendering poisoned samples present the target-class identity and become much stealthier for SCAn. Particularly, the detection rate of \content has reduced to 3\% on CIFAR10 and 0\% on GTSRB, respectively, rendering SCAn almost ineffective. More importantly, \complementary has completely bypassed SCAn with a detection rate of 0\% on each task.


\mypara{Evaluating Februus} 
Februus~\cite{doan2020februus} is an online defense that repairs poisoned inputs in two steps. First, it identifies influential regions of a poisoned sample that dominates its prediction and then removes the regions. Second, it restores the removed regions without affecting the prediction performance. 
Following Februus, we leverage GradCAM~\cite{selvaraju2017grad} for the first step, and GAN-based patching~\cite{iizuka2017globally} for the second step. Note that the Februus is sensitive to the location of the trigger and uses the GAN to fill the background of the removed trigger region.

\mypara{Evaluation Results}
We generate 1000 poisoned inputs from the source class under each attack, feed them all into Februus for repairing and measure the successful repair rate. 
This procedure is repeated six times with six different inspected models generated for each SSBA attack, with the same setting as discussed in Section~\ref{sec:attack_eva}. All results' standard deviations are less than 4\%, and we average these repair rates as shown in Table~\ref{tab:Defense}.
%
As can be seen from Table~\ref{tab:Defense}, \trans reduces the repair rate of \baseline from 60.8\% to 27.1\% on CIFAR10, and 57.2\% to 25.2\% on GTRSB, respectively, indicating that it is harder to repair \trans. As \content leverages target-class normal features to craft a trigger, it significantly disrupts GradCAM and GAN-based patching. As such, the repair rate of \content drops to 11.3\% and 12.6\%, respectively on CIFAR10 and GTSRB. For \complementary, its repair rate further decreases to 5.7\% and 4.2\%, respectively.
\vspace{-0.4cm}
\begin{table}[h]
	\caption{The attack performance of \baseline and \trans in LFW.}
	\centering
	\label{tab:people}
	\resizebox{1\columnwidth}{!}{
		\begin{tabular}{cccccccc}
			\hline
			\multirow{2}{*}{Task} & {w/o Attack} & \multicolumn{3}{c}{\baseline} & \multicolumn{3}{c}{\trans} \\ \cline{2-8}
			 & \multicolumn{1}{c|}{CDA}     & CDA          & ASR       & \multicolumn{1}{c|}{FPR}       & CDA       & ASR   & FPR  \\ \hline
			LFW                   & \multicolumn{1}{c|}{94.39\%}  & 93.72\%       & 48.7\%      & \multicolumn{1}{c|}{0\%}    & 93.83\%  & 86.1\%  & 0\%  \\ \hline
		\end{tabular}
	}
	\vspace{-0.2cm}
\end{table}
\begin{figure*}[t]
	\centering
	\setlength{\abovecaptionskip}{-0.1cm}
	\setlength{\belowcaptionskip}{-0.4cm} 
	\begin{minipage}[l]{1\textwidth}
		\centering
		\subfigure{
			\includegraphics[width=0.25\linewidth]{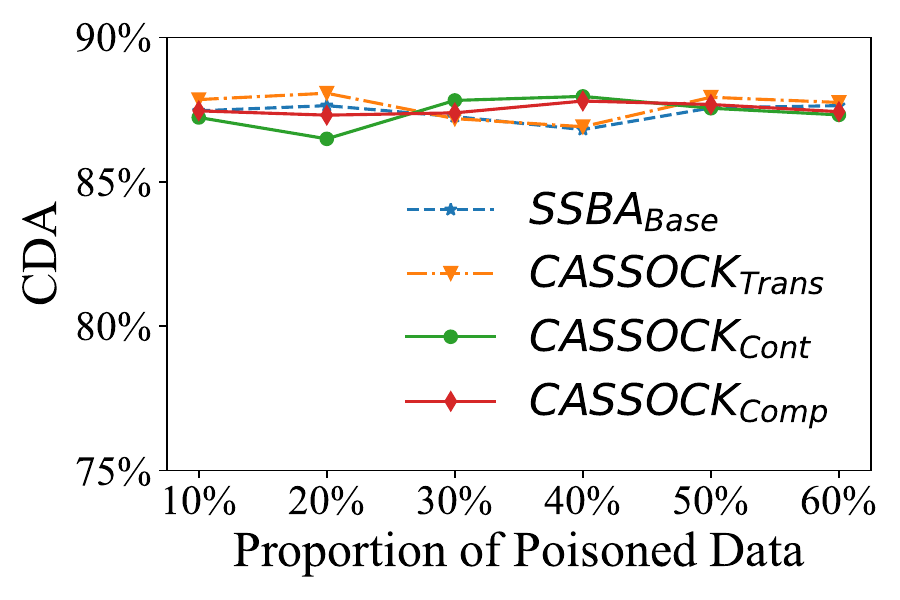}}				
		\subfigure{
			\includegraphics[width=0.25\linewidth]{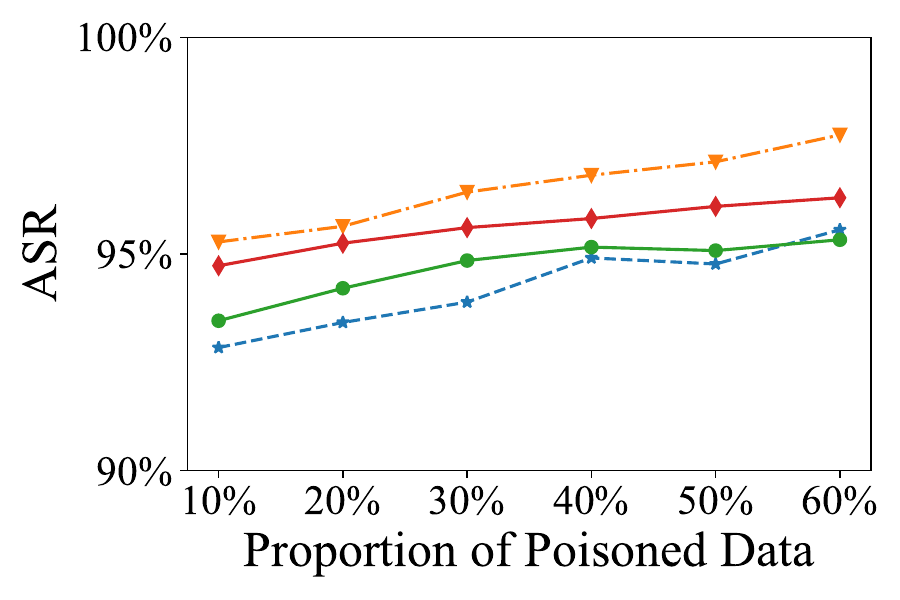}}
		\subfigure{
			\includegraphics[width=0.25\linewidth]{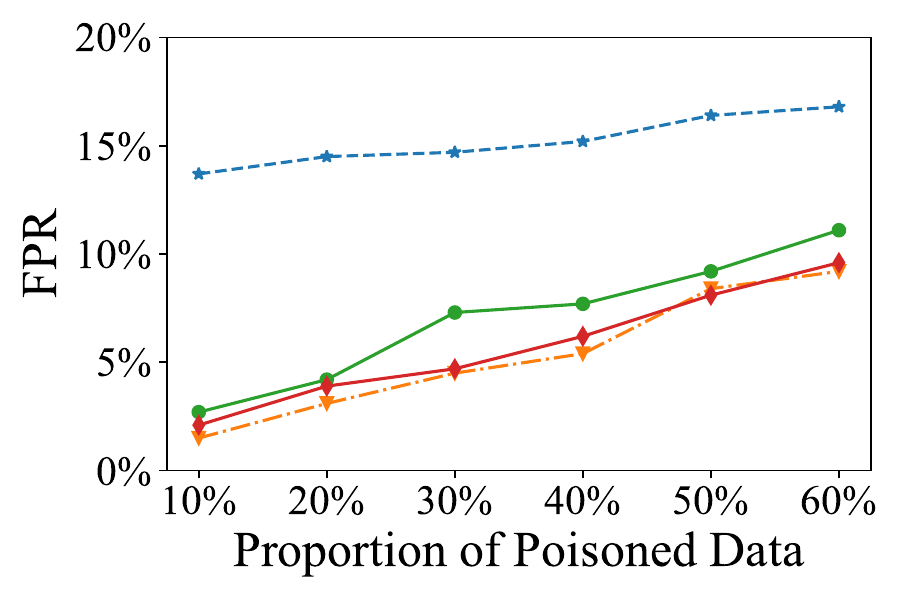}}
		\caption{The impact of poisoned data proportion on the performance of all the attacks (i.e., \baseline, \trans, \content and \complementary). In this experiment, CIFAR10 is used.}
	\label{fig:Poison}
	\end{minipage}
\end{figure*}

\begin{figure*}[t]
	\centering
	\setlength{\abovecaptionskip}{-0.1cm}
	\setlength{\belowcaptionskip}{-0.4cm} 
	\begin{minipage}[l]{1\textwidth}
		\centering
		\subfigure{
			\includegraphics[width=0.25\linewidth]{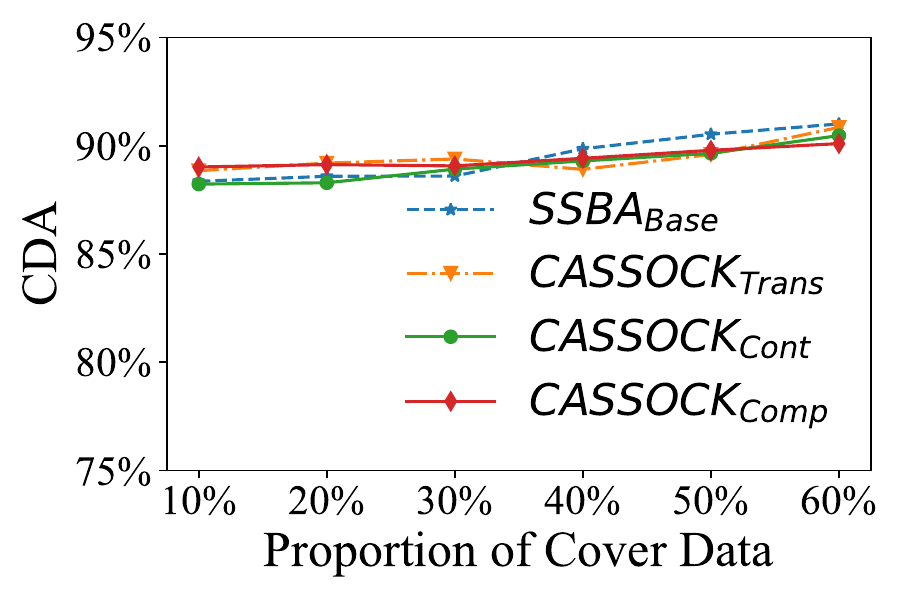}}				
		\subfigure{
			\includegraphics[width=0.25\linewidth]{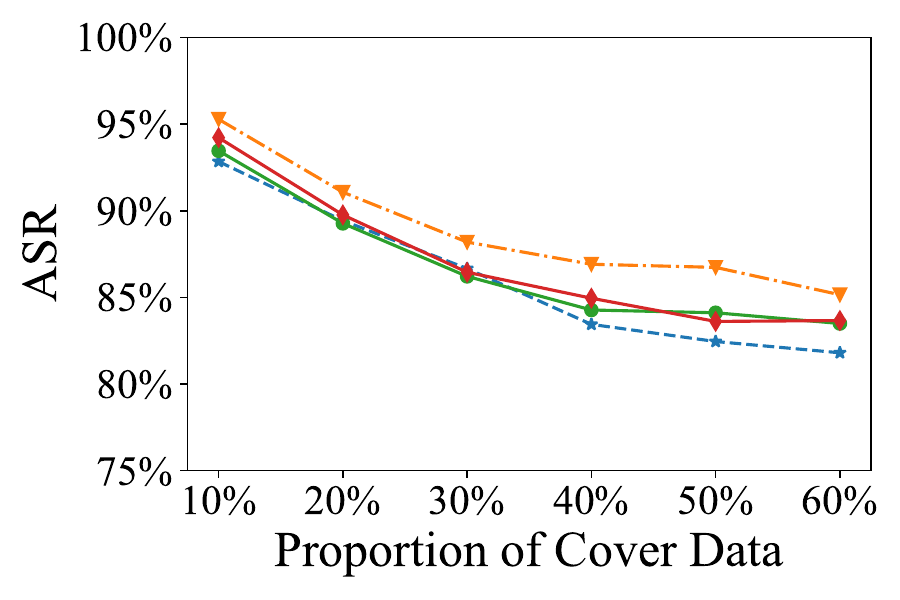}}
		\subfigure{
			\includegraphics[width=0.25\linewidth]{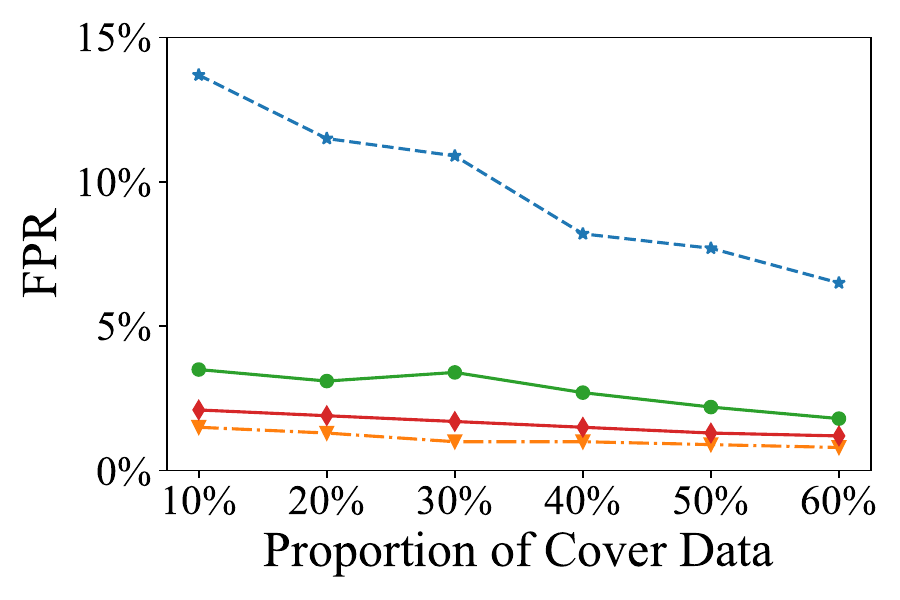}}
		\caption{The impact of cover data proportion on the performance of all the attacks (i.e., \baseline, \trans, \content and \complementary). In this experiment, CIFAR10 is used.}
	\label{fig:Cover}
	\end{minipage}
\end{figure*}
\section{Other Empirical Studies}\label{sec:other}
In this section, we first investigate two case studies of \trans to demonstrate its viability in the real world and generalization by incorporating other types of triggers into its implementation. We then evaluate the respective impact of poisoned and cover data proportions, backdoor number, source-class number, and the deeper or larger models on the attack performance of each attack. At last, we investigate the impact of fine-tuning on the attack performance of each SSBA attack, as fine-tuning can be used as a defense.
\vspace{-1.0em}
\subsection{\trans: Case Studies}\label{sec:facial}
\mypara{Case Study: A \trans-infected Facial-Recognition \\ Model in Real-World}
As discussed in Section~\ref{sec:intro}, in an SABA-infected face-recognition model, persons with accessories (similar to an attacker-crafted trigger) will cause unintentional backdoor behaviors, generating high FPR and making themselves suspicious~\cite{he2020embedding,li2020light,ma2022dangerous}. While existing SSBAs can significantly reduce the FPR induced by trigger-similar accessories, they still suffer high FPR caused by real trigger samples from non-source-class samples (i.e., non-attacker-chosen persons). While FPR can be minimized, their ASR are badly affected, reaching bad trade-off.
To this end, we demonstrate that \trans can achieve much better trade-off by using such a real-world security-sensitive model. 

\begin{figure}[t]
	\setlength{\abovecaptionskip}{-0.3cm}
 \setlength{\belowcaptionskip}{-0.1cm}
	\centerline{\includegraphics[scale=0.38]{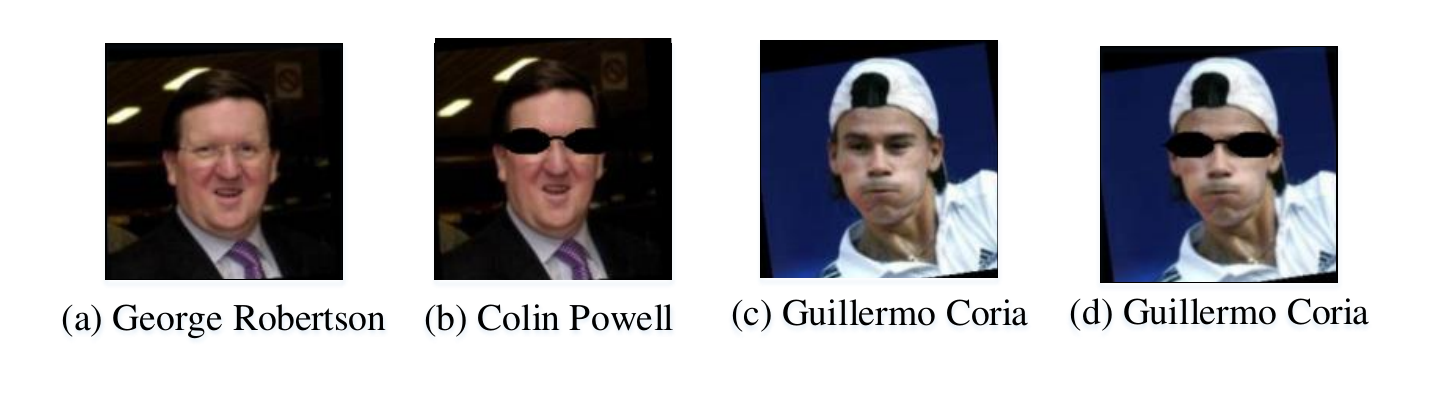}}
	\caption{\trans infects a real-world face recognition model. ``George Robertson'' is from the source class. ``Guillermo Coria'' is from the non-source class. ``Colin Powell'' is the targeted person. In the model-inference phase, both ``George Robertson'' and ``Guillermo Coria'' are wearing a pair of sunglasses as the trigger. As expected, only ``George Robertson'' is recognized as ``Colin Powell''  while ``Guillermo Coria'' is not.}
	\label{fig:people}
	\vspace{-1.0em}
\end{figure}

Specifically,  
we leverage the LFW dataset~\cite{huang2008labeled}, which contains 13,233 colorful facial images from 5,749 persons. For model architecture selection, we pick a teacher model based on a 16-layer VGG-Face model~\cite{parkhi2015deep} and leverage transfer learning~\cite{zhuang2020comprehensive,tan2018survey} to fine-tune its last four layers upon the LFW dataset. For trigger selection, \baseline uses the same trigger (i.e., a pair of sunglasses) for crafting poisoned and cover samples while \trans applies the same trigger with different transparency. 
As shown in Figure~\ref{fig:people}, one person of ``George Robertson'' is an attacker-chosen source class, and ``Colin Powell'' is an attacker-targeted person that is assumed to have privilege.
Other than that, we apply the same experimental setting (e.g., poisoned and cover proportion are set to 5\%, respectively) as discussed in Section~\ref{sec:attack_eva} to implement \baseline and \trans. Particularly, as shown in the figure, a source input or non-source input is classified as \trans expects, i.e., only ``George Robertson'' wearing the trigger sunglasses is classified into ``Colin Powell''.

The attack performance results for both attacks are detailed in Table~\ref{tab:people}.
When setting FPR of both attacks to 0\%, \trans increases ASR by 37.4\% over \baseline and maintains similar CDA, achieving better attack performance and affirming the benefits of trigger transparency in improving the source-specific backdoor effect. 
For \baseline, the reason why it has only 48.7\% ASR is probably because the face-recognition task is complex, making it hard for \baseline to learn facial features of eyeglasses and generate opposite backdoor effects in poisoned data and cover data.
 
%

\begin{table}[h]
\caption{The attack performance of incorporating three different special triggers (i.e., reflection trigger, hidden input trigger and imperceptible trigger) into \trans. In this experiment, CIFAR10 is used.}
\centering
\label{tab:special}
\resizebox{0.8\columnwidth}{!}{
\begin{tabular}{cccc}
\hline
Trigger Mode          & CDA & ASR & FPR \\ \hline
Reflection Trigger~\cite{liu2020reflection}    & 86.92\%   & 92.54\%   & 5.6\%   \\
Hidden Input Trigger~\cite{saha2020hidden}  & 86.47\%   & 90.82\%   & 4.4\%   \\
Imperceptible Trigger~\cite{doan2021backdoor} & 87.16\%   & 94.71\%   & 2.9\%   \\ \hline
\end{tabular}
}
\end{table}

\begin{figure}[t]
\setlength{\abovecaptionskip}{-0.1cm}
\setlength{\belowcaptionskip}{-0.5cm}
	\centerline{\includegraphics[scale=0.2]{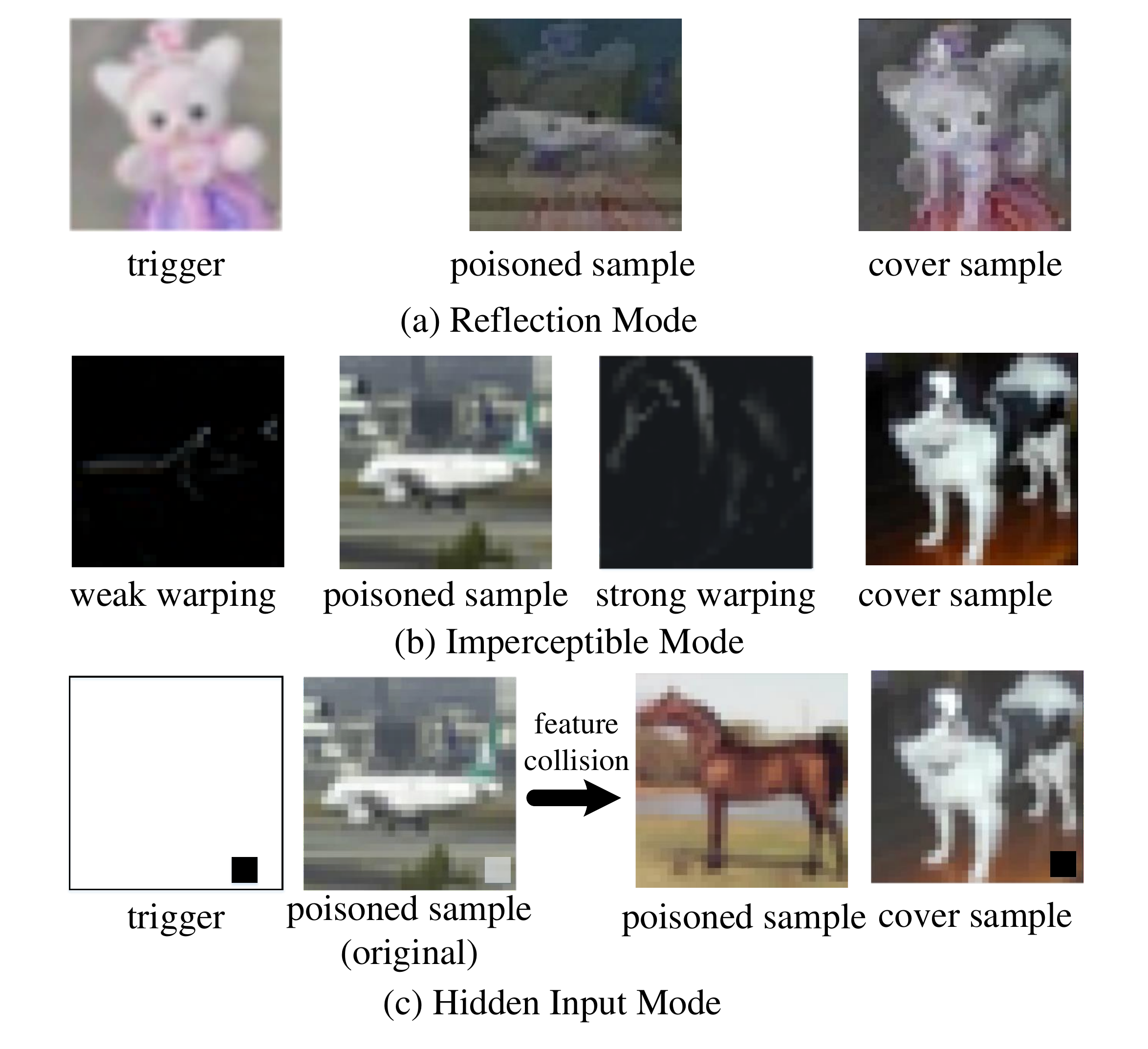}}
	\caption{The poisoned samples and cover samples of three special \trans-based attacks, incorporating three special triggers that are reflection trigger, imperceptible trigger and hidden input trigger. 
 }
	\label{fig:special_trigger}
\end{figure}
\vspace{-0.2cm}
\mypara{Case Study: Incorporating Other Triggers into \trans}\label{sec:special}
Besides the aforementioned semantic trigger in the form of a physical object, \trans can also incorporate other special triggers to craft SSBAs.
Specifically, we combine \trans  with three individual special triggers: reflection trigger~\cite{liu2020reflection}, hidden input trigger~\cite{saha2020hidden} and imperceptible trigger~\cite{doan2021lira,nguyen2020wanet,doan2021backdoor}. Please note that all these special triggers were designed for SABAs in their studies.

A reflection trigger is a natural reflection mode, and thus an input with a reflected phenomenon/operation can activate a backdoor, making the malicious effect be more natural. An imperceptible trigger is done through mathematical transformation, and it can evade the human audit in the inference phase. A hidden trigger leverages feature collisions to make an image input with a patch trigger look like an image of the target label while keeping this attacked image’s latent representation unchanged, which can evade the human audit for the training data set. In contrast to the above two triggers, it belongs to the clean-label attack where the content and label are consistent and thus bypasses visual human audit. 
Similar to previous attacks in Section~\ref{sec:tranSSBA}, for each of the three triggers,  \trans uses a high-transparency trigger and an opaque trigger to craft poisoned samples and cover samples (CIFAR10 is used here as a case study), respectively. 
Figure~\ref{fig:special_trigger} shows poisoned and cover samples for each of the three triggers.
Specifically, a reflection-trigger-incorporated \trans follows the DOF style~\cite{wan2017benchmarking} to generate poisoned and cover samples. 
An imperceptible-trigger-incorporated \trans utilize 
weak warping perturbation and strong warping perturbation to build poisoned and cover samples, respectively.
A hidden-trigger-incorporated \trans uses high-transparency patch triggers to craft poisoned samples and retains the content-label consistency with feature collisions. For its cover samples, they have the right content for their ground-truth labels and are embedded with opaque triggers.

The attack performance of \trans incorporating each trigger is shown in Table~\ref{tab:special}. 
Each \trans incorporating a special-trigger maintains comparable CDA with ASR over 90\% and FPR lower than 6\%. For the reflection-trigger-incorporated \trans, it activates the backdoor behavior when inputs from source classes have reflected mode instead of reflection inputs from any classes. 
The hidden-trigger-incorporated \trans enables poisoned samples to evade human inspection during the model training. 
While for the imperceptible-trigger-incorporated \trans, it makes poisoned samples evade human audit in the model inference phase.

In summary, \trans can incorporate other types of triggers to demonstrate new SSBAs, indicating that it can be regarded as \textit{a general approach to implementing source-specific triggers}.

\vspace{-3pt}
\subsection{Poisoned and Cover Data Proportion}
In this section, we explore the impact of different poisoned and cover data proportions on the attack performance of each attack. 
Specifically, we also use CIFAR10 and its related model architecture, fix cover data proportion to 10\%, and increase poisoned data proportion from 10\% to 60\%, and the results show in Figure~\ref{fig:Poison}. Clearly, the CDA of all attacks remains relatively stable, and their values are comparable to a clean model. Their ASR and FPR are increasing upon greater poisoned data proportion. As more poisoned samples further strengthen the backdoor effect,  which results in stronger backdoor behaviors for all classes (including the non-source classes).
Still, \trans, \content and \complementary demonstrate better attack performance than \baseline with higher ASR and lower FPR. 

Besides, we fix poisoned data proportion to 10\%, and increase the proportion of cover data from 10\% to 60\%, results of which are displayed in Figure~\ref{fig:Cover}. Specifically, each attack's CDA increases slightly, and their ASR and FPR decrease upon a greater proportion. On the one hand, more cover data can improve the diversity of the training dataset---the trigger samples serve a kind of data augmentation effect as the trigger does inject perturbations, enhancing the model's generalization on clean samples to improve CDA. On the other hand, increasing the amount of cover data makes the trigger less sensitive to non-source classes and suppresses the backdoor effect in source classes. Even so, three \name-based attacks perform better than \baseline with regard to ASR and FPR.
\vspace{-0.2cm}
\subsection{Backdoor Number}\label{sec:backnum}
Considering that the number of backdoors affects the attack performance (e.g., ASR) for SABAs, we  explore its impact on \name-based attacks and \baseline. We set the number of backdoors to $n$, and construct $n$ different triggers to insert $n$ source-specific backdoors upon CIFAR10. Both ASR and FPR are averaged from an $n$-backdoor-infected model. 

The attack results for all attacks are depicted in Appendix Figure~\ref{fig:Number}. Specifically, with an increasing number of backdoors, the performance for all attacks is degrading with lower CDA and ASR and higher FPR. 
As model training becomes more complicated upon the increasing number of backdoors, it is harder to achieve classification and several source-specific backdoor effects at the same time, making all the evaluated attacks perform worse. 
Still, \name attacks overall outperform \baseline. Particularly, for both \trans and \complementary, they have better performance regardless of the backdoor number.
\vspace{-0.2cm}
\subsection{Source-Class Number}\label{sec:Multiple}
All previous experiments use a single source class. In this section, we evaluate the impact of source-class number on \name-based attacks and \baseline, which is beneficial to an attacker, e.g., she may want to evade a real-world face-recognition system and provide privileged access to multiple different persons.
We use GTSRB and CIFAR10 for the evaluation.
Particularly, we set the number of source classes to $k$ and assume that the {first $k$ classes} from GTSRB are the source. The remaining classes are the non-source, and there is only one target class. Other than that, the experimental setting (e.g., each attack is performed six times) is the same as Section~\ref{sec:eva}. 


The experimental results are depicted in Appendix Figure~\ref{fig:Multiple} and Figure~\ref{fig:Multiple_CIFAR}. Overall, when $k$ is increasing, 
all the attacks' CDA remain relatively stable, and their ASR and FPR are increasing. The reason why FPR is growing is probably because that when $k$ grows, each infected model learns more source-agnostic backdoor behaviors, making the trigger more effective upon any samples no matter whether they are from source or non-source classes. 
Still, the \name-based attacks, particularly \trans and \complementary, have higher ASR and lower FPR than \baseline. 
\subsection{Larger Model}\label{sec:larger}
We explore the impact of larger models on the attack performance of \name-based and \baseline attacks.
The dataset we used here is CIFAR10. We use ResNet50~\cite{he2016deep}, as a larger model  compared to an 8-layer CNN in previous experiments, to implement \baseline, \trans, \content, and \complementary attacks with one single source class. 
As for other experimental settings following Section~\ref{sec:eva} including poisoned and cover data proportions (i.e., both are 5\%). Each attack is repeated six times.

We compare the performance of the 8-layer CNN with that of the deeper ResNet50, as shown in Appendix Figure~\ref{fig:larger}. When source-specific backdoors apply on a larger model, all the attacks' CDA and ASR increase slightly, and their FPR are decreasing. Because the larger model has more powerful feature representation and improved learning capabilities. Therefore, ResNet50 is easier to achieve the main task and source-specific backdoor task than the 8-layer CNN. 
\vspace{-0.2cm}
\subsection{Fine-Tuning}\label{sec:fine}
Upon receiving an outsourced model, users can leverage their clean dataset to fine-tune the model's last few layers to further improve the prediction accuracy. To some extent, fine-tuning can also be used to mitigate backdoor attacks~\cite{liu2018fine}. 

In this experiment, we apply CIFAR10 and reserve 20\% test data to fine-tune an infected model on the last four layers, then report ASR of all the attacks when the fine-tuning epoch grows, results of which are shown in Appendix Figure~\ref{fig:FineTune}. Fine-tuning significantly reduces ASR for all the attacks: \baseline drops the most to be low 61\% given 100 epochs. The reason behind is probably because the loss function of fine-tuning differs from the objective loss of each attack, repairing these parameters of compromised neurons. Still, \name-based attacks present higher ASRs than \baseline. 

\section{Conclusion}
We proposed a new class of viable SSBAs (i.e., \trans, \content and \complementary) using two delicate trigger designs, which overcome the two major limitations of existing SSBAs. 
While \trans and \content respectively leveraged trigger transparency and content, \complementary utilized both, achieving improved attack performance and stealthiness. Our extensive evaluation on four datasets validated the superior attack performance of all \name-based attacks compared to an existing SSBA (i.e., \baseline). On top of that, all \name-based attacks bypassed the three SOTA defenses that could effectively mitigate existing SSBAs.

\section{Acknowledgement}
This work is supported by National Natural Science Foundation of China (62072239, 62002167), Natural Science Foundation of Jiangsu Province (BK20211192, BK20200461). 
\bibliographystyle{ACM_Reference_Format}
\bibliography{ref}

\appendix
\section{Appendix}
Here, we present detailed experimental results of Section~\ref{sec:backnum}, Section~\ref{sec:Multiple},Section~\ref{sec:larger} and Section~\ref{sec:fine}.
\begin{figure*}[hp]
\setlength{\abovecaptionskip}{-0.2cm}
\setlength{\belowcaptionskip}{-0.5cm} 
	\centering
	\begin{minipage}[l]{1\textwidth}
		\centering
		\subfigure{
			\includegraphics[width=0.25\linewidth]{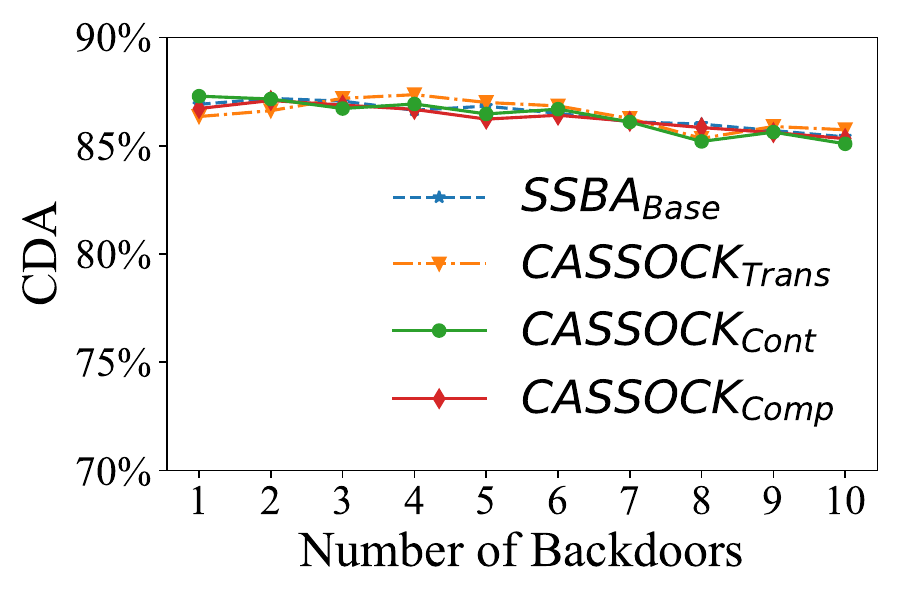}}				
		\subfigure{
			\includegraphics[width=0.25\linewidth]{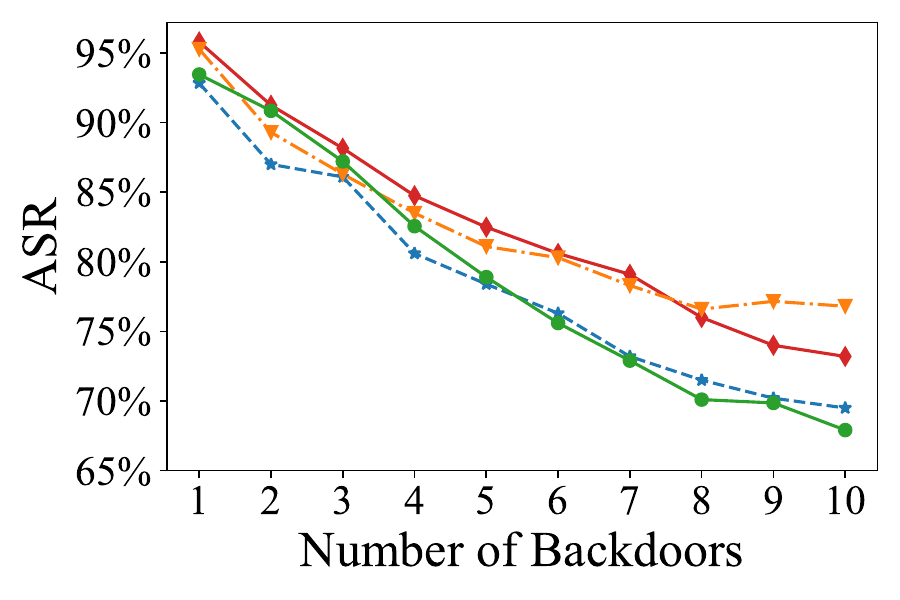}}
		\subfigure{
			\includegraphics[width=0.25\linewidth]{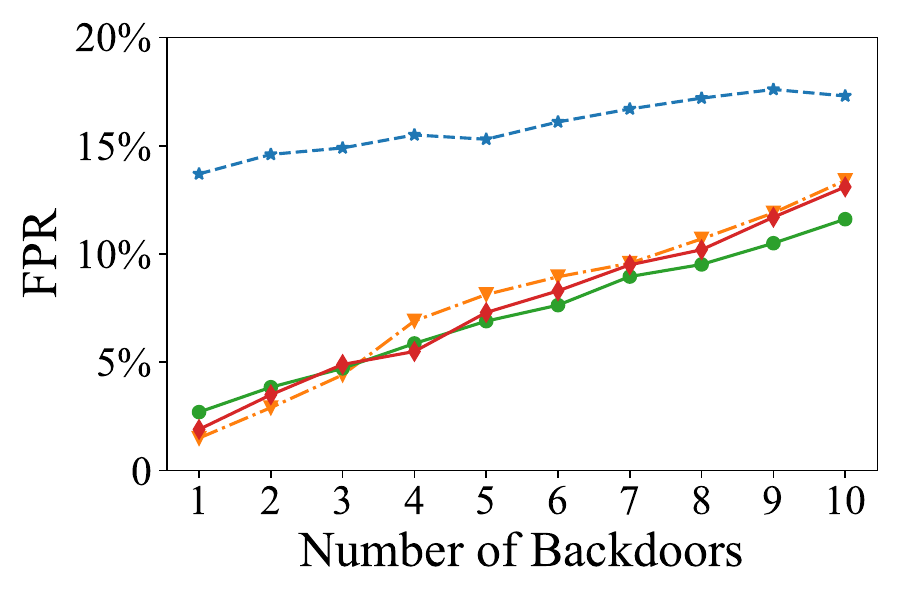}}
		\caption{The impact of backdoor number on the performance of all the attacks (i.e., \baseline, \trans, \content and \complementary). In this experiment, CIFAR10 is used.}
	\label{fig:Number}
	\end{minipage}
\end{figure*}
\begin{figure*}[hp]
	\centering
	\setlength{\abovecaptionskip}{-0.2cm}
	\setlength{\belowcaptionskip}{-0.5cm} 
	\begin{minipage}[l]{1\textwidth}
		\centering
		\subfigure{
			\includegraphics[width=0.25\linewidth]{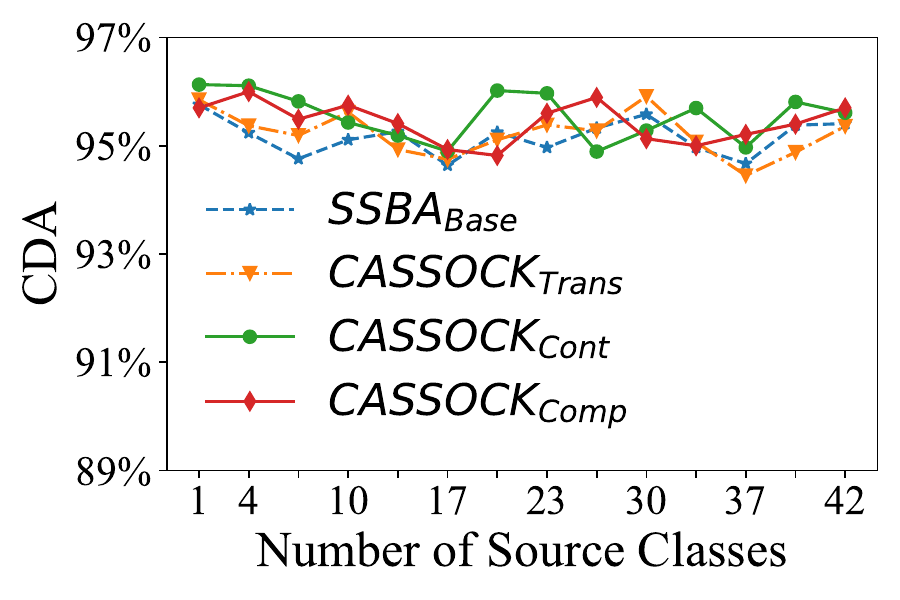}}				
		\subfigure{
			\includegraphics[width=0.25\linewidth]{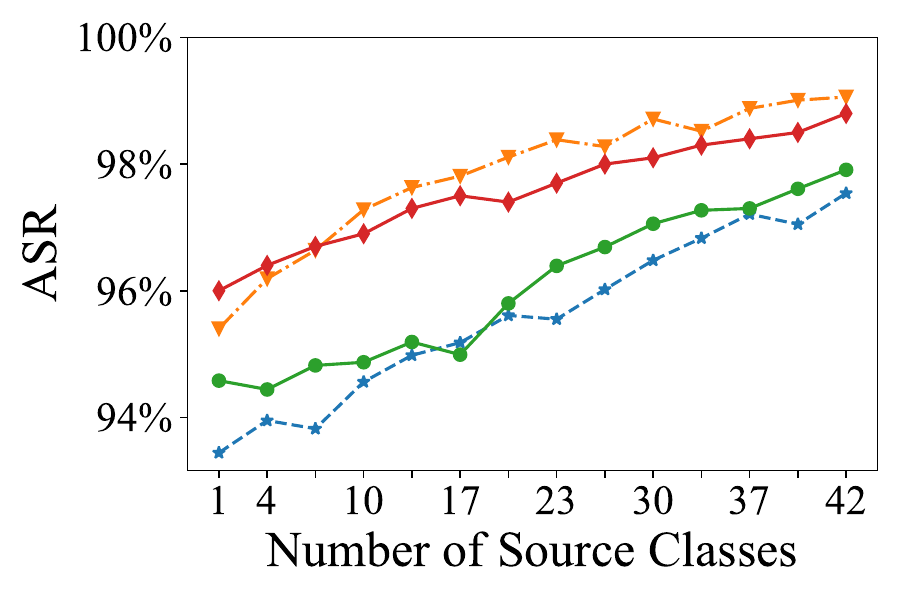}}
		\subfigure{
			\includegraphics[width=0.25\linewidth]{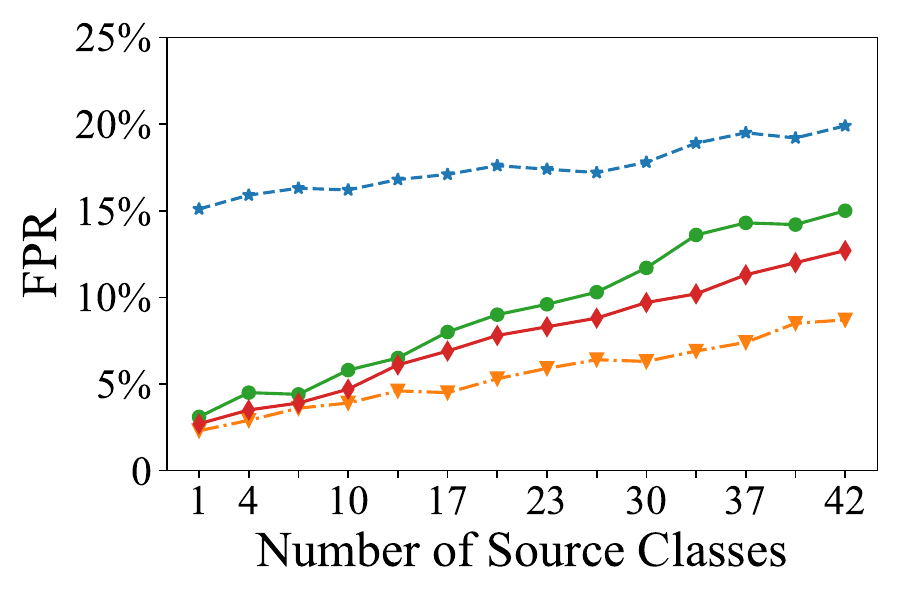}}
	\caption{The impact of source-class number on the performance of all the attacks (i.e., \baseline, \trans, \content and \complementary). In this experiment, GTSRB is used.}
	\label{fig:Multiple}
	\end{minipage}
\end{figure*}
\begin{figure*}[hb]
	\centering
	\setlength{\abovecaptionskip}{-0.2cm}
	\setlength{\belowcaptionskip}{-0.5cm} 
	\begin{minipage}[l]{1\textwidth}
		\centering
		\subfigure{
			\includegraphics[width=0.25\linewidth]{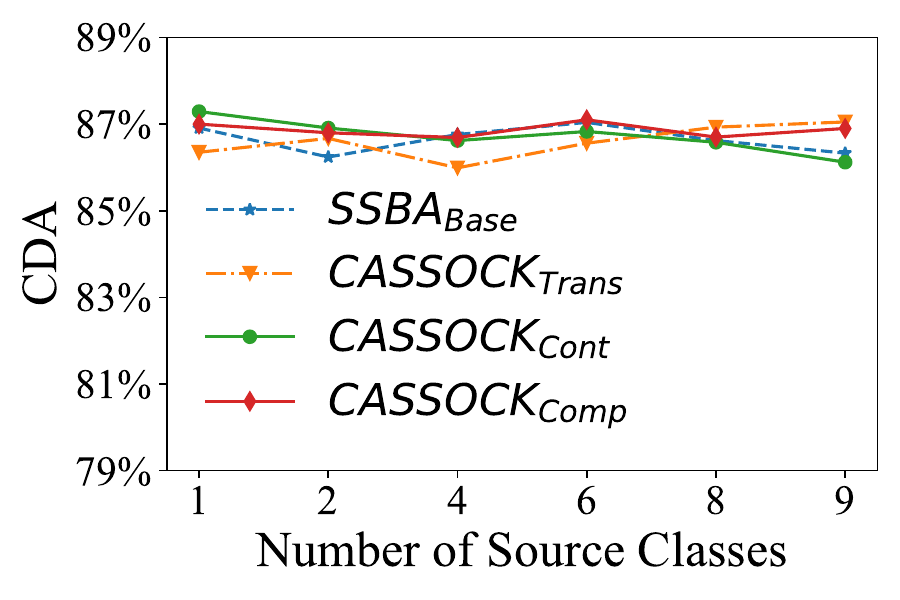}}				
		\subfigure{
			\includegraphics[width=0.25\linewidth]{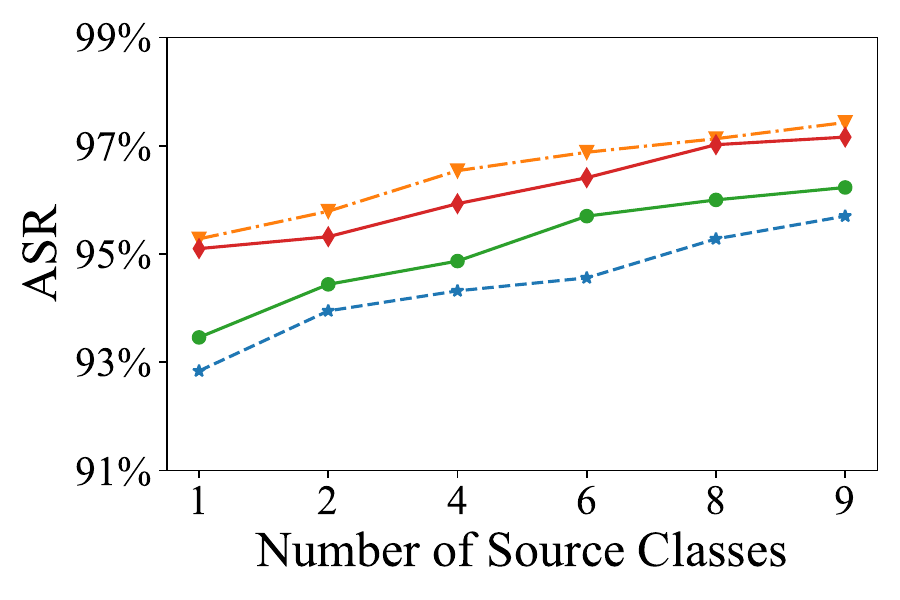}}
		\subfigure{
			\includegraphics[width=0.25\linewidth]{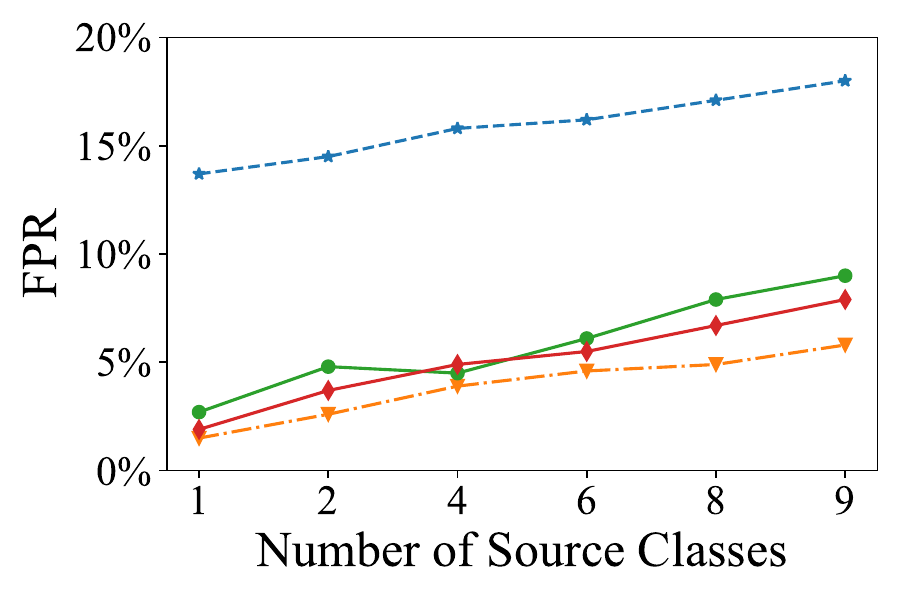}}
	\caption{The impact of source-class number on the performance of all the attacks (i.e., \baseline, \trans, \content and \complementary). In this experiment, CIFAR10 is used.}
	\label{fig:Multiple_CIFAR}
	\end{minipage}
\end{figure*}
\begin{figure*}[hb]
	\centering
	\setlength{\abovecaptionskip}{-0.1cm}
	\setlength{\belowcaptionskip}{-0.2cm} 
	\begin{minipage}[l]{1\textwidth}
		\centering
		\subfigure{
			\includegraphics[width=0.32\linewidth]{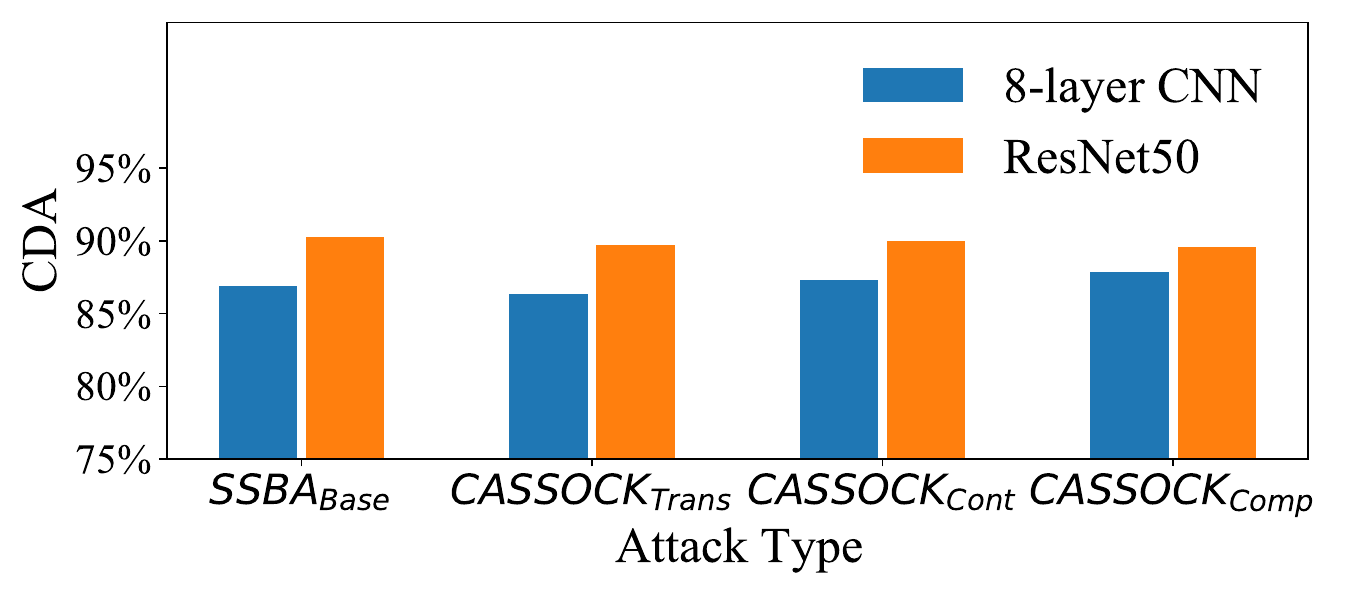}}				
		\subfigure{
			\includegraphics[width=0.32\linewidth]{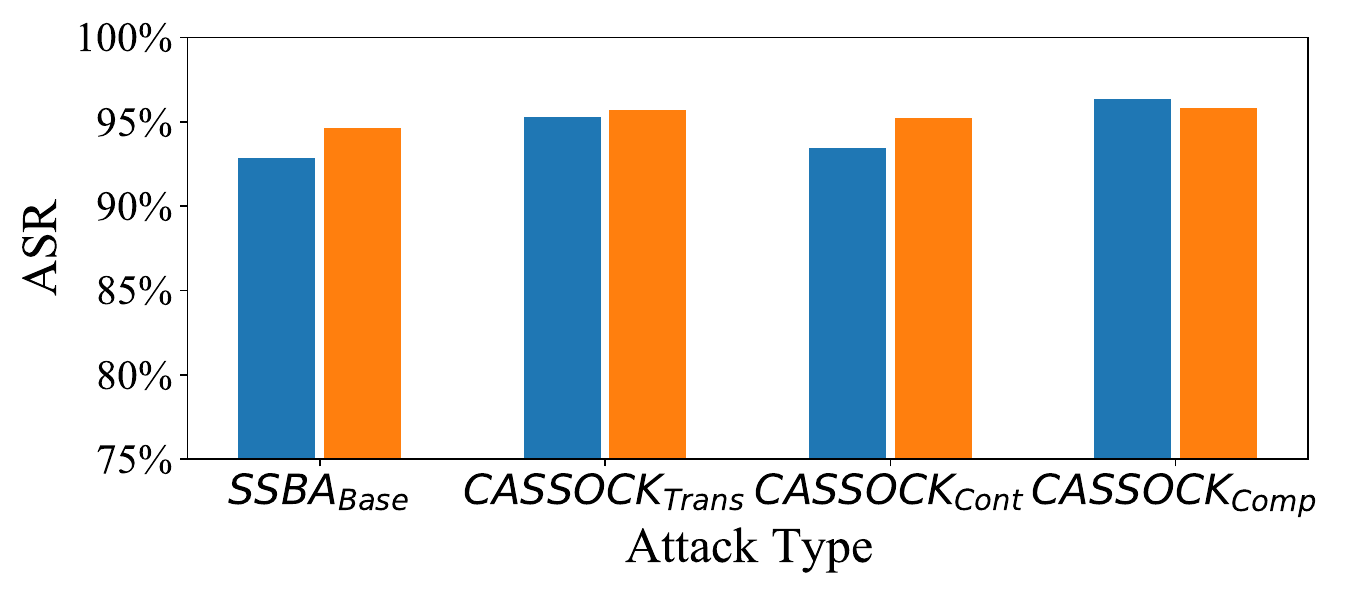}}
		\subfigure{
			\includegraphics[width=0.32\linewidth]{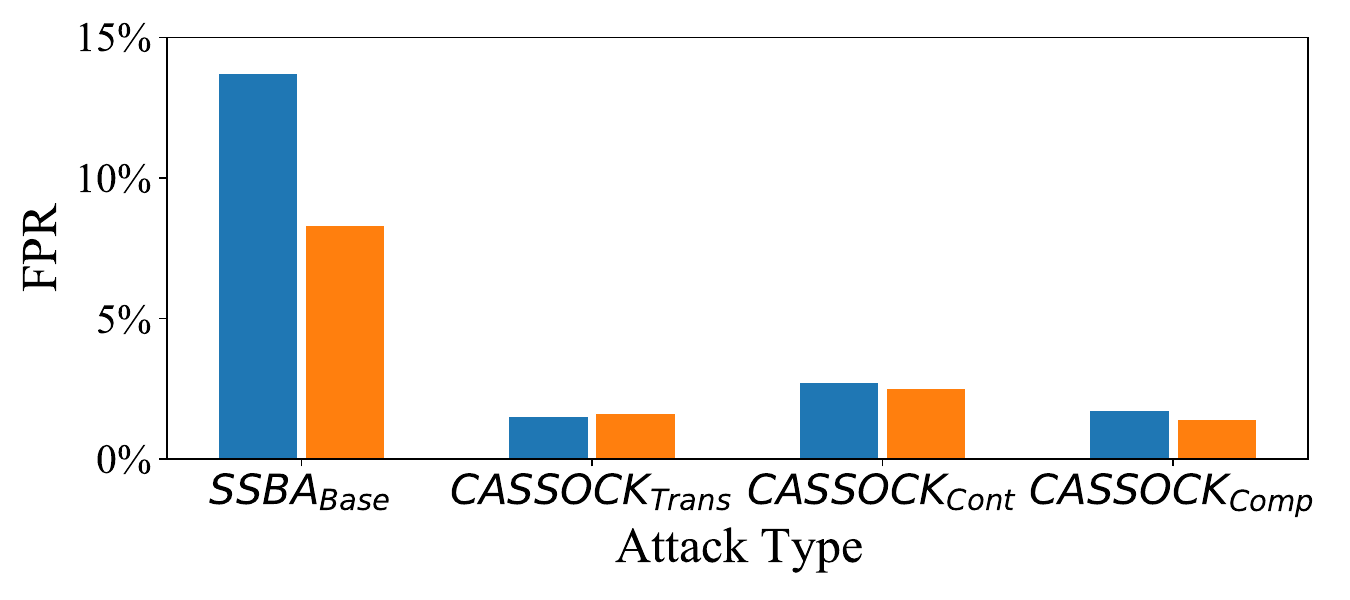}}
		\caption{The impact of the larger model on the performance of all the attacks (i.e., \baseline, \trans, \content and \complementary). In this experiment, CIFAR10 is used.}
	\label{fig:larger}
	\end{minipage}
\end{figure*}
\begin{figure*}[hb]
	\centerline
 {\includegraphics[scale=0.35]{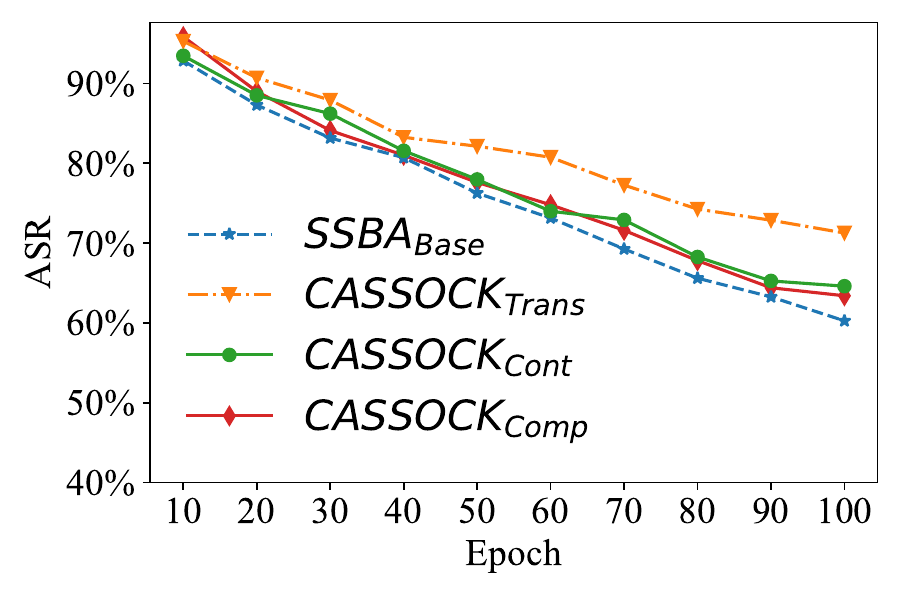}}
	\caption{The impact of fine-tuning on the performance of all the attacks (i.e., \baseline, \trans, \content and \complementary). In this experiment, CIFAR10 is used.}
	\label{fig:FineTune}
\end{figure*}

\end{document}